\documentclass[twocolumn,prd,superscriptaddress,altaffilletter,nofootinbib,prep,preprintnumbers]{revtex4-1}
\usepackage{amsmath}
\usepackage{amssymb}
\usepackage{amsfonts}
\usepackage{comment}
\usepackage{graphicx}
\usepackage[colorlinks=true,citecolor=blue,urlcolor=blue,breaklinks=true]{hyperref}
\usepackage{acronym}
\usepackage[usenames,dvipsnames]{xcolor}
\usepackage{xspace}
\usepackage{tikz}
\usepackage{siunitx}
\usepackage{makecell}
\usepackage{multirow}
\usepackage{bm}
\usepackage{romannum}
\usepackage{xcolor}
\usepackage{orcidlink}
\usepackage[caption=false]{subfig}
\usepackage{ulem}
\usepackage{booktabs}
\usepackage{physics}
\usepackage{tensor}
\usepackage{braket}

\usetikzlibrary{arrows,shapes,trees,decorations.pathreplacing}
\allowdisplaybreaks

\newcommand{\gw}[0]{\ac{GW}\xspace}
\newcommand{\gws}[0]{\acp{GW}\xspace}
\newcommand{\bh}[0]{\ac{BH}\xspace}
\newcommand{\bbh}[0]{\ac{BBH}\xspace}
\newcommand{\bbhs}[0]{\acp{BBH}\xspace}
\newcommand{\qnm}[0]{\ac{QNM}\xspace}
\newcommand{\qnms}[0]{\acp{QNM}\xspace}
\newcommand{\gr}[0]{\ac{GR}\xspace}
\newcommand{\mcmc}[0]{\ac{MCMC}\xspace}
\newcommand{\snr}[0]{\ac{SNR}\xspace}
\newcommand{\snrs}[0]{\acp{SNR}\xspace}
\newcommand{\sxs}[0]{\ac{SXS}\xspace}

\newcommand{\psd}[0]{\ac{PSD}\xspace}
\newcommand{\hpd}[0]{\ac{HPD}\xspace}
\newcommand{\othree}[0]{\ac{O3}\xspace}
\newcommand{\ofour}[0]{\ac{O4}\xspace}
\newcommand{\ofive}[0]{\ac{O5}\xspace}

\newcommand{\e}{\mathrm{e}}
\newcommand{\iu}{\mathrm{i}}

\begin{document}
\preprint{RESCEU-16/25}
\pagenumbering{arabic}

\title{Analyzing black-hole ringdowns with orthonormal modes}

\author{Soichiro Morisaki \orcidlink{0000-0002-8445-6747}}
\affiliation{Institute for Cosmic Ray Research, The University of Tokyo, 
5-1-5 Kashiwanoha, Kashiwa, Chiba 277-8582, Japan}
\author{Hayato Motohashi \orcidlink{0000-0002-4330-7024}}
\affiliation{Department of Physics, Tokyo Metropolitan University, 1-1 Minami-Osawa, Hachioji, Tokyo 192-0397, Japan}
\author{Motoki Suzuki \orcidlink{0009-0009-3585-0762}}
\affiliation{Department of Physics, Graduate School of Science, The University of Tokyo,
7-3-1 Hongo, Bunkyo-ku, Tokyo 113-8655, Japan}
\affiliation{Institute for Cosmic Ray Research, The University of Tokyo, 
5-1-5 Kashiwanoha, Kashiwa, Chiba 277-8582, Japan}
\author{Daiki Watarai
\orcidlink{0009-0002-7569-5823}}
\affiliation{Department of Physics, Graduate School of Science, The University of Tokyo,
7-3-1 Hongo, Bunkyo-ku, Tokyo 113-8655, Japan}
\affiliation{Research Center for the Early Universe (RESCEU), Graduate School of Science, The University of Tokyo, Tokyo 113-0033, Japan}

\begin{abstract}

The ringdown signal following a \bh merger can be modeled as a superposition of \bh \qnms, offering a clean setup for testing gravitational theories.
In particular, detecting multiple \qnms enables consistency checks of their frequencies and damping times, serving as a test of general relativity---a technique known as black hole spectroscopy.
However, incorporating additional \qnms introduces challenges such as increased parameter correlations and higher computational costs in data analysis.
To address this, we propose an efficient Bayesian analysis method that applies the Gram-Schmidt algorithm to the \qnms.
This reduces the correlation between the modes and enables analytic marginalization over the mode amplitudes.
We validate our approach using damped sinusoids and numerical waveforms from the Simulating eXtreme Spacetimes catalog.

\end{abstract}

\maketitle

\acrodef{GW}{gravitational wave}
\acrodef{BBH}{binary black hole}
\acrodef{BH}{black hole}
\acrodef{QNM}{quasinormal mode}
\acrodef{GR}{general relativity}
\acrodef{NR}{Numerical Relativity}
\acrodef{MCMC}{Markov chain Monte Carlo}
\acrodef{SNR}{signal-to-noise ratio}
\acrodef{SXS}{Simulating eXtreme Spacetimes}
\acrodef{KDE}{kernel density estimation}
\acrodef{PSD}{power spectral density}
\acrodef{HPD}{highest posterior density}
\acrodef{O3}{third observing run}
\acrodef{O4}{fourth observing run}
\acrodef{O5}{fifth observing run}

\acresetall 

\section{Introduction}

Since the first direct detection of \gws, the number of detections has increased rapidly over the past decade.
The LIGO-Virgo-KAGRA collaboration reported 90 \gw signals observed until the end of its \othree \cite{LIGOScientific:2018mvr, LIGOScientific:2020ibl, LIGOScientific:2021usb, KAGRA:2021vkt}, with hundreds of additional significant event candidates identified in the ongoing \ofour.
Most of these signals are consistent with \gws emitted by mergers of \bbhs.
These detections provide valuable insights into the properties of gravity in the strong-field regime \cite{LIGOScientific:2016lio,LIGOScientific:2018dkp,LIGOScientific:2019fpa,LIGOScientific:2020ibl,KAGRA:2021vkt}.

In particular, the ringdown part of GW signal, which is emitted while the merger remnant is settling down to a single rotating black hole, provides a clean setting for testing gravitational theories. 
At the linear order, it can be modeled as a superposition of damped sinusoids, referred to as \qnms.
Each mode is labeled by a set of integers, $(l,m,n)$, where $(l,m)$ are two angular numbers and $n$ is an overtone index that orders modes by decreasing damping time.
In \gr, the frequency and damping time of each \qnm depend solely on the mass and spin of the remnant black hole, allowing these parameters to be estimated from \qnm measurements \cite{Echeverria:1989hg}.
A simple test of \gr is to compare the remnant mass and spin inferred from the ringdown with those independently estimated from the premerger part to check their consistency \cite{Hughes:2004vw, Ghosh:2016qgn, Ghosh:2017gfp}.

If multiple \qnms are observed, an alternative approach is to check the consistency of their frequencies and damping times, known as \bh spectroscopy \cite{Dreyer:2003bv, Berti:2005ys, Berti:2007zu, Berti:2025hly}.
This technique offers a relatively clean test of \gr,
as it does not rely on detailed waveform modeling.
Detecting modes beyond the dominant $(l,|m|, n) = (2, 2, 0)$ mode, however, requires a high signal-to-noise ratio. 
Nevertheless, evidence for additional modes has been reported in several events \cite{Isi:2019aib, Capano:2021etf,Siegel:2023lxl,LIGOScientific:2020ibl,KAGRA:2021vkt, Gennari_2024}.

A notable case is GW150914 \cite{LIGOScientific:2016aoc}.
Isi et al.\ analyzed GW150914 incorporating overtones ($n>0$) and found evidence of at least one overtone for $l=m=2$ with $3.6\sigma$ confidence \cite{Isi:2019aib}.
While they fit the data starting from the signal peak, where the \qnm approximation may not be strictly valid, their analysis, including overtones, yields mass and spin estimations consistent with those from the full inspiral-merger-ringdown analysis.
Furthermore, their \bh spectroscopy analysis found the observed \qnms to be consistent with the predictions of \gr.
Subsequent studies by other groups have revisited these findings, with some reporting reduced evidence for overtones compared to the original analysis \cite{Wang:2021elt, Cotesta:2022pci, Finch:2022ynt, Ma:2023cwe, Correia:2023bfn, Wang:2024yhb}.
The validity of fitting data from the signal peak using \qnms, particularly regarding the stability of the fitting, has also been the subject of discussion in the literature \cite{Baibhav:2023clw, Nee:2023osy, Cheung:2023vki, Takahashi:2023tkb, Clarke:2024lwi, Giesler:2024hcr, Gao:2025zvl, Mitman:2025hgy, Giesler:2019uxc, Bhagwat_2020, Finch_2021, Dhani_2021, Forteza_2021}.

For accurate inference, it is desirable to include as many \qnms as possible.
However, increasing the number of included modes also increases the number of model parameters, making the analysis more complex and computationally expensive. 
Moreover, since \qnms are not orthogonal, correlations between modes make it challenging to unambiguously identify which modes are present in the data, especially when many modes are involved.

In this paper, we propose an efficient method for ringdown inference involving multiple \qnms.
Specifically, we employ the Gram-Schmidt procedure to orthonormalize the \qnm basis and sample the mode coefficients using a prior that is uniform in their amplitudes.
This parametrization and prior choice offer two key advantages.
First, the mode coefficients can be analytically marginalized over, substantially reducing the dimensionality of the inference problem and lowering computational cost.
Second, orthonormalization mitigates correlations between mode coefficients, enhancing the robustness of mode identification.

We note that several authors have previously proposed alternative methods for analytically marginalizing over mode amplitudes \cite{Ma:2023cwe,Dong:2025igh}.
The rational filter technique in \cite{Ma:2023cwe} is employed to remove the dominant mode, and in combination with random sampling methods such as \mcmc, it forms a hybrid approach to obtain the posteriors of the subdominant modes.
In contrast, our approach follows standard Bayesian inference and allows us to obtain the full posterior distributions over all parameters without relying on random sampling methods.
This enables a direct understanding of the correlations among the parameters.
Both our method and the FIREFLY algorithm in \cite{Dong:2025igh} allow for analytic marginalization, but our approach differs in that it makes use of orthonormal modes.
Furthermore, the application of the Gram-Schmidt algorithm to the \qnm basis was explored in a different context in \cite{Sago:2021gbq}.

This paper is organized as follows.
Section~\ref{sec:review} reviews the ringdown waveform model and time-domain Bayesian analysis framework.
In Sec.~\ref{sec:method}, we introduce our analysis method.
Sections~\ref{sec:damped} and~\ref{sec:sxs} present validation of our method using both damped sinusoids and numerical relativity waveforms from the \sxs catalog.
Finally, Sec.~\ref{sec:conclusion} summarizes our conclusions.

\section{Time-domain Bayesian framework for analyzing ringdown} \label{sec:review}

In this Section, we introduce the signal model we employ in this work and review a time-domain Bayesian method to infer signal parameters.

\subsection{Signal model}

Let $h(t)$ denote complex \gw strain,
\begin{equation}
    h(t) = h_+(t) + \iu h_\times(t),
\end{equation}
where $h_+(t)$ and $h_\times(t)$ are plus and cross polarizations of 
\gws, respectively.
The ringdown signal is modeled as a superposition of damped sinusoids labeled by a set of indices $\bm{\alpha}=(l, m, n)$ \cite{Berti:2005ys,Isi:2021iql},
\begin{equation}
    \begin{aligned}
        h(t) &= \sum_{\bm{\alpha}} \Big[ \mathcal{C}_{\bm{\alpha}} \e^{-\iu \tilde{\omega}_{\bm{\alpha}} t} S_{\bm{\alpha}} (M_f\chi_f\tilde{\omega}_{\bm{\alpha}}; \iota, \phi) \\
        &\hspace{5em} +\mathcal{C}'_{\bm{\alpha}} \e^{\iu \tilde{\omega}^\ast_{\bm{\alpha}} t} S_{\bm{\alpha}'} (M_f\chi_f\tilde{\omega}_{\bm{\alpha}}; \iota, \phi)\Big], \label{eq:polarization}
    \end{aligned}
\end{equation}
where $\mathcal{C}_{\bm{\alpha}}$ and $\mathcal{C}'_{\bm{\alpha}}$ are complex mode amplitudes, $\tilde{\omega}_{\bm{\alpha}} = \omega_{\bm{\alpha}} - \iu / \tau_{\bm{\alpha}}$ is the complex frequency dependent on the remnant \bh's mass and spin, $M_f$ and $\chi_f$.
$S_{\bm{\alpha}} (M_f\chi_f\tilde{\omega}_{\bm{\alpha}};\iota, \phi)$ is the spin-weighted spheroidal harmonics with the spin weight of $-2$, dependent on the mass $M_f$, spin $\chi_f$, and polar and azimuthal angles $\iota$ and $\phi$, and $\bm{\alpha}'=(l,-m,n)$.
The summation runs over $\bm{\alpha}=(l, m, n)$ with $\sum_{\bm{\alpha}} = \sum_{l=2}^\infty \sum_{m=-l}^l \sum_{n=0}^\infty$.
Here, the summation does not group \qnms based on their angular dependence, but is instead reorganized so that the terms with $m>0$ correspond to prograde modes, while those with $m<0$ correspond to retrograde modes.
The overtone index $n$ orders \qnms for a given $l$ and $m$ by decreasing decay time, i.e., $\tau_{lm(n+1)} < \tau_{lmn}$, at $\chi_f=0$.

Let $I$ be the index of the detector.
The gravitational wave strain observed in the $I$th detector is given by
\begin{equation}
    h^I(t) = \Re \left[F^I h(t - t^I_{\mathrm{S}})\right], \label{eq:hI}
\end{equation}  
where $F^I = F^{I}_+ - \iu F^{I}_{\times}$
represents the complex beam pattern of the $I$th detector and $t^I_{\mathrm{S}}$ is the analysis start time 
at the $I$th detector.
Substituting Eq.~\eqref{eq:polarization} into Eq.~\eqref{eq:hI}, we obtain 
\begin{equation}
    h^I(t) = \sum_{\bm{\alpha}} \sum_{j=0}^3 c_{j, \bm{\alpha}} v^I_{j, \bm{\alpha}}(t), \label{eq:hI2}
\end{equation}
where
\begin{align}
    &c_{0, \bm{\alpha}} = \Re \left[\mathcal{C}_{\bm{\alpha}} S_{\bm{\alpha}} (\iota, \phi) + \mathcal{C}'_{\bm{\alpha}} S_{\bm{\alpha}'} (\iota, \phi)\right], \\
    &c_{1, \bm{\alpha}} = \Im \left[\mathcal{C}_{\bm{\alpha}} S_{\bm{\alpha}} (\iota, \phi) - \mathcal{C}'_{\bm{\alpha}} S_{\bm{\alpha}'} (\iota, \phi)\right], \\
    &c_{2, \bm{\alpha}} = \Im \left[\mathcal{C}_{\bm{\alpha}} S_{\bm{\alpha}} (\iota, \phi) + \mathcal{C}'_{\bm{\alpha}} S_{\bm{\alpha}'} (\iota, \phi)\right], \\
    &c_{3, \bm{\alpha}} = \Re \left[- \mathcal{C}_{\bm{\alpha}} S_{\bm{\alpha}} (\iota, \phi) + \mathcal{C}'_{\bm{\alpha}} S_{\bm{\alpha}'} (\iota, \phi)\right],
\end{align}
and
\begin{align}
    v^I_{0, \bm{\alpha}}(t) &= F^I_{+} \e^{-\frac{t - t^I_{\mathrm{S}}}{\tau_{\bm{\alpha}}}} \cos (\omega_{\bm{\alpha}} (t - t^I_{\mathrm{S}})), \label{eq:template basis0} \\
    v^I_{1, \bm{\alpha}}(t) &= F^I_{+} \e^{-\frac{t - t^I_{\mathrm{S}}}{\tau_{\bm{\alpha}}}} \sin (\omega_{\bm{\alpha}} (t - t^I_{\mathrm{S}})), \\
    v^I_{2, \bm{\alpha}}(t) &= F^I_{\times} \e^{-\frac{t - t^I_{\mathrm{S}}}{\tau_{\bm{\alpha}}}} \cos (\omega_{\bm{\alpha}}(t - t^I_{\mathrm{S}})), \\
    v^I_{3, \bm{\alpha}}(t) &= F^I_{\times} \e^{-\frac{t - t^I_{\mathrm{S}}}{\tau_{\bm{\alpha}}}} \sin (\omega_{\bm{\alpha}} (t - t^I_{\mathrm{S}})).
    \label{eq:template basis3}
\end{align}

Throughout this work, we assume the coefficients $c_{j,\bm{\alpha}}$ are free parameters that can take arbitrary real values, without imposing any specific amplitude model.
The angular dependence on $\iota$ and $\phi$ is fully absorbed into $\{c_{j,\bm{\alpha}}\}$.
Similarly, although the beam pattern $F^I$ depends on the polarization angle $\psi$ via $F^I(\psi) = F^I(\psi=0) \e^{2 \iu \psi}$, this dependence is also absorbed into $\{c_{j,\bm{\alpha}}\}$.
Following previous works \cite{Isi:2019aib,LIGOScientific:2020ibl,KAGRA:2021vkt}, we fix the sky location, $\alpha$ and $\delta$, whose values are typically determined based on the information provided by the inspiral-merger-ringdown analysis.
Thus, the beam pattern $F^I$ is treated as a fixed quantity.

The analysis start time $t^I_{\mathrm{S}}$ is also fixed.
Typically, the start time at a reference detector is fixed near the signal peak time, and the start times at the other detectors are determined by the time delays computed from the fixed sky location.
Consequently, the signal model is fully characterized by the remnant \bh's mass and spin, $M_f$ and $\chi_f$, and the set of real coefficients $\{c_{j,\bm{\alpha}}\}$.

In real analyses, only a finite number of \qnms are considered, under the assumption that contributions from the others are negligible.
Let $K$ denote the number of included \qnms, indexed as $\{\bm{\alpha}_0,\bm{\alpha}_1,\dots,\bm{\alpha}_{K-1}\}$.
In this notation, we assume that \qnms are ordered by significance---i.e., modes with $\bm{\alpha}_{k+1}$ are less significant than those with $\bm{\alpha}_{k}$.
In our demonstration presented in later Sections, $\alpha_0$ represents the dominant $(2, 2, 0)$, while $\alpha_k$ with $k>0$ denotes the overtones for $l=m=2$, $\alpha_k=(2, 2, k)$. 

When data are available from a single detector, or when detectors are nearly co-aligned, $v^I_{2, \bm{\alpha}}(t)$ and $v^I_{3, \bm{\alpha}}(t)$ become degenerate with $v^I_{0, \bm{\alpha}}(t)$ and $v^I_{1, \bm{\alpha}}(t)$.
In such cases, we set $c_{2, \bm{\alpha}}=c_{3, \bm{\alpha}}=0$, retaining only two coefficients per mode.
Let $D=2$ or $4$ denote the number of coefficients used per mode.
The resulting model becomes,
\begin{equation}
    h^I(t) = \sum_{k=0}^{K-1} \sum_{j=0}^{D-1} c_{j, \bm{\alpha}_k} v^I_{j, \bm{\alpha}_k}(t). \label{eq:hI3}
\end{equation}
Thus, the set of real parameters in the analysis consists of $K \times D$ real coefficients:
\begin{equation}
    \{c_{j,\bm{\alpha}_k}|j=0,\dots,D-1;k=0,\dots,K-1\}. \nonumber
\end{equation}
For brevity, we refer to this set as $\{c_{j,\bm{\alpha}_k}\}$ throughout the text.

It is worth noting that $c_{j, \bm{\alpha}}$ corresponds to $x_{+/\times}$ and $y_{+/\times}$ in \cite{Isi:2021iql}.
The relation between this and other parametrizations of \gw polarizations is discussed in \cite{Isi:2022mbx}.

\subsection{Bayesian inference with time-domain likelihood}

Next, we review a Bayesian method to infer the signal parameters, $\bm{\theta}=\{M_f,\chi_f\} \cup \{c_{j,\bm{\alpha}_k}\}$.
The posterior probability density function of $\bm{\theta}$ is computed via Bayes's theorem,
\begin{equation}
    p(\bm{\theta}) = \frac{\pi(\bm{\theta}) \mathcal{L}(\bm{\theta})}{\mathcal{Z}},
\end{equation}
where $\pi(\bm{\theta})$ is the prior probability density function, $\mathcal{L}(\bm{\theta})$ is the likelihood function, and $\mathcal{Z}$ is the evidence, serving as a normalization constant.

Following \cite{Isi:2021iql}, we employ a time-domain likelihood function given by,
\begin{equation}
    \begin{aligned}
        &\ln \mathcal{L}(\bm{\theta}) = \\
        &~~ - \frac{1}{2} \sum_I \left(\bm{d}^I - \bm{h}^I(\bm{\theta})\right)^T \left(R^I\right)^{-1} \left(\bm{d}^I - \bm{h}^I(\bm{\theta})\right).
    \end{aligned}
\end{equation}
Here, $\bm{d}^I$ and $\bm{h}^I(\bm{\theta})$ represent vectors consisting of data and waveform values, respectively, for the $I$th detector:
\begin{equation}
    \bm{d}^I \equiv \begin{pmatrix}
        d^I(t^I_0) \\
        d^I(t^I_1) \\
        d^I(t^I_2) \\
        \vdots
    \end{pmatrix},~~~\bm{h}^I(\bm{\theta}) \equiv \begin{pmatrix}
        h^I(t^I_0; \bm{\theta}) \\
        h^I(t^I_1; \bm{\theta}) \\
        h^I(t^I_2; \bm{\theta}) \\
        \vdots
    \end{pmatrix},
\end{equation}
where $t^I_{i}$ represents the $i$th time sample of the $I$th detector.
The components of the matrix $R^I$ are given by the noise autocorrelation function evaluated at the time lag between samples,
\begin{equation}
    \left[R^I\right]_{ij} = R^I(|t_{i}^I - t_{j}^I|).
\end{equation}

As in many previous studies, we assume that the prior is separable,
\begin{equation}
    \pi(\bm{\theta}) = \pi(M_f, \chi_f) \pi(\{c_{j,\bm{\alpha}_k}\}).
\end{equation}
Typically, $\pi(M_f, \chi_f)$ is taken to be uniform over a broad range of $M_f$ and $\chi_f$.
On the other hand, several different forms of $\pi(\{c_{j,\bm{\alpha}_k}\})$ have been proposed and employed in the literature \cite{Isi:2021iql,Isi:2022mbx,LIGOScientific:2020ibl,Capano:2021etf}.

\section{Ringdown analysis with orthonormal modes} \label{sec:method}

In this Section, we introduce our semianalytic analysis method. 
As a preparatory step, we first rewrite the likelihood function.
We define concatenated data and waveform vectors across all detectors,
\begin{equation}
    \bm{d} \equiv \begin{pmatrix}
        \bm{d}^{I_0} \\
        \bm{d}^{I_1} \\
        \bm{d}^{I_2} \\
        \vdots 
    \end{pmatrix},~~~\bm{h}(\bm{\theta}) \equiv \begin{pmatrix}
        \bm{h}^{I_0}(\bm{\theta}) \\
        \bm{h}^{I_1}(\bm{\theta}) \\
        \bm{h}^{I_2}(\bm{\theta}) \\
        \vdots
    \end{pmatrix}.
\end{equation}
We also introduce the block-diagonal noise correlation matrix,
\begin{equation}
    R = \begin{pmatrix}
        R^{I_0} & 0 & 0 & \cdots \\
        0 & R^{I_1} & 0 & \cdots \\
        0 & 0 & R^{I_2} & \cdots \\
        \vdots & \vdots & \vdots & \ddots
    \end{pmatrix}.
\end{equation}
With these variables, the logarithm of likelihood takes the following form,
\begin{align}
    \ln \mathcal{L}(\bm{\theta}) &= - \frac{1}{2} \left(\bm{d} - \bm{h}(\bm{\theta})\right)^T R^{-1} \left(\bm{d} - \bm{h}(\bm{\theta})\right) \\
    &= \bm{d}^T R^{-1} \bm{h}(\bm{\theta}) - \frac{1}{2} \bm{h}^T(\bm{\theta}) R^{-1} \bm{h}(\bm{\theta}) + \mathrm{const.} \label{eq:likelihood1}
\end{align}

Let $\bm{v}^I_{j, \bm{\alpha}}$ be a vector consisting of $v^I_{j, \bm{\alpha}}(t)$ evaluated at the time samples $\{t^I_i\}$,
\begin{equation}
    \bm{v}^I_{j, \bm{\alpha}} = \begin{pmatrix}
        v^{I}_{j, \bm{\alpha}}(t^I_0) \\
        v^{I}_{j, \bm{\alpha}}(t^I_1) \\
        v^{I}_{j, \bm{\alpha}}(t^I_2) \\
        \vdots \\
    \end{pmatrix}.
\end{equation}
We then define the concatenated vector over all detectors,
\begin{equation}
    \bm{v}_{j,\bm{\alpha}} = \begin{pmatrix}
        \bm{v}^{I_0}_{j, \bm{\alpha}} \\
        \bm{v}^{I_1}_{j, \bm{\alpha}} \\
        \bm{v}^{I_2}_{j, \bm{\alpha}} \\
        \vdots
    \end{pmatrix}.
\end{equation}
Equation~\eqref{eq:hI3} implies that the concatenated waveform vector $\bm{h}(\bm{\theta})$ can be expressed as a linear combination of these basis vectors,
\begin{equation}
    \bm{h}(\bm{\theta}) = \sum_{k=0}^{K-1} \sum_{j=0}^{D - 1} c_{j, \bm{\alpha}_k} \bm{v}_{j, \bm{\alpha}_k} = V \bm{c}, \label{eq:template}
\end{equation}
where
\begin{align}
    V &= \left(\bm{v}_{0, \bm{\alpha}_0}, \ldots, \bm{v}_{D - 1, \bm{\alpha}_0}, \bm{v}_{0, \bm{\alpha}_1}, \ldots, \bm{v}_{D-1, \bm{\alpha}_{K-1}}\right), \\
    \bm{c} &= \left(c_{0, \bm{\alpha}_0}, \ldots, c_{D - 1, \bm{\alpha}_0}, c_{0, \bm{\alpha}_1}, \ldots, c_{D-1, \bm{\alpha}_{K-1}}\right)^T.
\end{align}
Substituting Eq.~\eqref{eq:template} into Eq.~\eqref{eq:likelihood1}, we obtain
\begin{equation}
    \ln \mathcal{L}(\bm{\theta}) = \bm{d}^T R^{-1} V \bm{c} - \frac{1}{2} \bm{c}^T V^T R^{-1} V \bm{c} + \mathrm{const.} \label{eq:likelihood2}
\end{equation}

\subsection{Gram-Schmidt orthogonalization} \label{sec:gs}

We perform Gram-Schmidt orthogonalization of the basis vectors $\bm{v}_{j, \bm{\alpha}_k}$ with respect to the inner product defined by,
\begin{equation}\label{eq:inner product}
    \left(\bm{v}_{j,\bm{\alpha}_k}, \bm{v}_{j',\bm{\alpha}_{k'}}\right) = \bm{v}_{j,\bm{\alpha}_k}^T R^{-1} \bm{v}_{j',\bm{\alpha}_{k'}}.
\end{equation}
This procedure yields an orthonormal set of basis vectors $\tilde{\bm{v}}_{j,\bm{\alpha}_k}$, which we assemble into the matrix,
\begin{align}
    \tilde{V} &= \left(\tilde{\bm{v}}_{0, \bm{\alpha}_0}, \ldots, \tilde{\bm{v}}_{D - 1, \bm{\alpha}_0}, \tilde{\bm{v}}_{0, \bm{\alpha}_1}, \ldots \tilde{\bm{v}}_{D-1, \bm{\alpha}_{K-1}}\right) \\
    &= V U, \label{eq:def_U}
\end{align}
where $U$ is an upper triangular matrix obtained from the orthogonalization process.
Note that we continue to label the orthonormal vectors using the original indices $(j, \bm{\alpha}_k)$.

\begin{figure}[h]
    \centering
    \includegraphics[width=\linewidth]{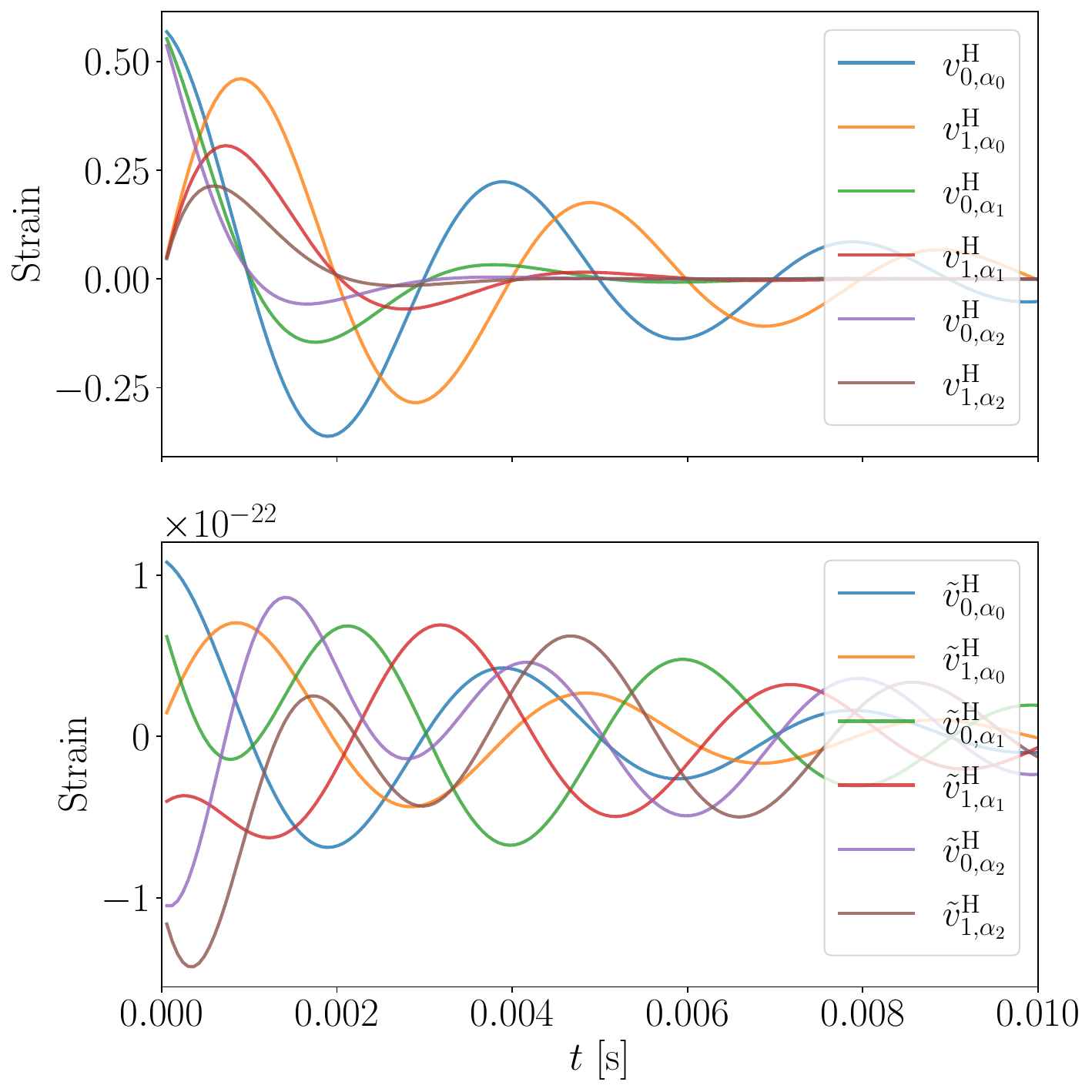}
    \caption{
        Top panel shows the template basis composed of the original (nonorthogonal) modes $v_{j,\alpha_k}^I(t)$, as presented in Eqs.~\eqref{eq:template basis0}--\eqref{eq:template basis3}.
        We fix $M_f=68.2\,M_{\odot}$, $\chi_f=0.692$, and $t^I_{\mathrm{S}}=0$.
        The bottom panel shows the orthonormal template basis $\tilde{v}^I_{j,\alpha_k}(t)$ obtained by orthogonalizing the original basis.
        Here, we adopt the noise \psd estimated with the data around GW150914, available from \cite{LIGO-P1900011-v1}.
        We focus solely on the $l=m=2$ modes with $D=2$ and include up to the second overtone, fixing the detector $I$ to Hanford.
    }
    \label{fig:comparison template waveforms}
\end{figure}

The top and bottom panels of Fig.~\ref{fig:comparison template waveforms}, respectively, show the template basis $v^I_{j,\bm{\alpha}_k}(t)$ presented in Eqs.~\eqref{eq:template basis0}--\eqref{eq:template basis3} and the orthonormal template basis $\tilde{v}^I_{j,\bm{\alpha}_k}(t)$ obtained through the orthogonalization procedure described above.
Compared to the original basis, the orthonormal basis functions appear more distinct from one another, indicating that they form a more independent set and are thus expected to provide a more efficient representation of ringdown signal.
Note that $\tilde{v}^I_{j,\bm{\alpha}_k}(t)$ scales with the noise level, as it is normalized using the inner product~\eqref{eq:inner product} defined by the noise correlation matrix $R$.

By construction, the new basis vectors satisfy
\begin{equation}
    \tilde{V}^T R^{-1} \tilde{V} = \mathbb{I}. \label{eq:orthonormality}
\end{equation}
In terms of the orthonormal basis, the likelihood takes the following form,
\begin{equation}
    \ln \mathcal{L}(\bm{\theta}) = \tilde{\bm{d}}^T \tilde{\bm{c}} - \frac{1}{2} \tilde{\bm{c}}^T \tilde{\bm{c}} + \mathrm{const.}, \label{eq:likelihood3}
\end{equation}
where
\begin{equation}
    \tilde{\bm{d}} = \tilde{V}^T R^{-1} \bm{d},~~~\tilde{\bm{c}} = U^{-1} \bm{c}. \label{eq:amplitude_transformation}
\end{equation}
This expression shows that the likelihood function is Gaussian in $\tilde{\bm{c}}$ with diagonal covariance.
In contrast, Eq.~\eqref{eq:likelihood2} shows that the likelihood is Gaussian in $\bm{c}$ with covariance of $\left(V^T R^{-1} V\right)^{-1}$, which generally contains off-diagonal terms.
Thus, the transformation reduces parameter correlations in the posterior.
This advantage is further demonstrated through simulation studies presented in later sections.

The orthonormal vectors for $\bm{\alpha}_k$ with $k>0$ span the subspace orthogonal to all basis vectors associated with earlier modes $\{\bm{\alpha}_{k'}\}_{k' < k}$.
As a result, they isolate components that cannot be represented by earlier modes but are present in the $\bm{\alpha}_k$ modes.
These orthonormal basis vectors are constructed through the Gram-Schmidt orthogonalization, which depends on the ordering of the modes being orthogonalized.
Assuming that the modes are orthogonalized in order of their significance, the mode amplitude defined by
\begin{align}
    \tilde{A}_{\bm{\alpha}_k} \equiv \sqrt{\sum_{j=0}^{D - 1} \left(\tilde{c}_{j, \bm{\alpha}_k}\right)^2},
    \label{eq:tilde_A}
\end{align}
is nonzero if the original mode amplitude
\begin{align}
    A_{\bm{\alpha}_k}\equiv\sqrt{\sum_{j=0}^{D-1}(c_{j,\bm{\alpha}_k})^2},
    \label{eq:A}
\end{align}
is nonzero.
Based on this consistency, we interpret the nonzero values of $\tilde{A}_{\bm{\alpha}_k}$ as evidence for the presence of the $\bm{\alpha}_k$ mode in the data.

Additionally, the input vectors for the Gram--Schmidt orthogonalization can be constructed as arbitrary linear combinations of the damped sinusoids given in Eqs.~\eqref{eq:template basis0}–\eqref{eq:template basis3}.
While the coefficients $\tilde{c}_{j,\bm{\alpha}_k}$ of the orthonormalized basis naturally vary depending on the choice of the input vectors, the mode amplitudes $\tilde{A}_{\bm{\alpha}_k}$ remain invariant.

\subsection{Marginalization over mode amplitudes}

One of the key advantages of using orthonormal basis vectors is that, under a prior that is uniform in the mode amplitudes, the posterior can be analytically marginalized over the mode coefficients.
With the mode amplitude $\tilde{A}_{\bm{\alpha}_k}$, the likelihood function is expressed as
\begin{equation}
    \begin{aligned}
        &\mathcal{L} \propto \prod_{k} \mathcal{L}_{\bm{\alpha}_k},\\
        &\mathcal{L}_{\bm{\alpha}_k} \equiv \exp \left[- \frac{1}{2} \tilde{A}^2_{\bm{\alpha}_k} + \tilde{A}_{\bm{\alpha}_k} \tilde{d}_{\bm{\alpha}_k} \cos \beta_{\bm{\alpha}_k}\right].
    \end{aligned} \label{eq:likelihood per mode}
\end{equation}
Here, $\tilde{d}_{\bm{\alpha}_k}$ and $\beta_{\bm{\alpha}_k}$ denote the $L^2$ norm of $\tilde{\bm{d}}_{\bm{\alpha}_k}$ and the angle between $\tilde{\bm{d}}_{\bm{\alpha}_k}$ and $\tilde{\bm{c}}_{\bm{\alpha}_k}$, respectively, where $\tilde{\bm{d}}_{\bm{\alpha}_k}$ and $\tilde{\bm{c}}_{\bm{\alpha}_k}$ are the subvectors of $\tilde{\bm{d}}$ and $\tilde{\bm{c}}$ collecting the components associated with the $\bm{\alpha}_k$ modes.
The prior uniform in the mode amplitudes is given by
\begin{equation}
    \pi(\{c_{j,\bm{\alpha}_k}\}) \prod_{j, k} d c_{j,\bm{\alpha}_k} \propto \prod_{k} \frac{1}{\tilde{A}_{\bm{\alpha}_k}^{D - 1}} \prod_{j, k} d \tilde{c}_{j,\bm{\alpha}_k}.
    \label{eq:amplitude prior}
\end{equation}

With this prior, the likelihood for a single mode, $\mathcal{L}_{\bm{\alpha}}$, can be analytically marginalized over the mode coefficients.
The likelihood marginalized over the coefficients with the mode amplitude fixed is proportional to
\begin{align}
    \mathcal{L}^{\beta}_{\bm{\alpha}} &\equiv \int_0^\pi \sin^{D - 2} \beta_{\bm{\alpha}} d \beta_{\bm{\alpha}} \mathcal{L}_{\bm{\alpha}} \nonumber \\
    &\begin{aligned}
        &= 2^{\frac{D}{2} - 1} \sqrt{\pi} \Gamma\left(\frac{D-1}{2}\right) \times \\
        &~~ (\tilde{A}_{\bm{\alpha}} \tilde{d}_{\bm{\alpha}})^{-\frac{D}{2}} \e^{-\frac{\tilde{A}_{\bm{\alpha}}^2}{2}} \left(D I_{\frac{D}{2}} (\tilde{A}_{\bm{\alpha}} \tilde{d}_{\bm{\alpha}}) + \tilde{A}_{\bm{\alpha}} \tilde{d}_{\bm{\alpha}} I_{\frac{D + 2}{2}} (\tilde{A}_{\bm{\alpha}} \tilde{d}_{\bm{\alpha}})\right).
    \end{aligned} \nonumber \\
    &= \begin{cases}
    \displaystyle \pi \e^{-\frac{\tilde{A}_{\bm{\alpha}}^2}{2}} I_{0} (\tilde{A}_{\bm{\alpha}} \tilde{d}_{\bm{\alpha}}) & (D=2), \\
    \displaystyle \pi \e^{-\frac{\tilde{A}_{\bm{\alpha}}^2}{2}} (\tilde{A}_{\bm{\alpha}} \tilde{d}_{\bm{\alpha}})^{-1} I_{1} (\tilde{A}_{\bm{\alpha}} \tilde{d}_{\bm{\alpha}}) & (D=4),
    \end{cases} \label{eq:L_beta}
\end{align}
where $I_n(z)$ is the modified Bessel function of the first kind.
Finally, marginalizing over the mode amplitude yields
\begin{align}
    \mathcal{L}^{A, \beta}_{\bm{\alpha}} &\equiv \int_0^\infty d \tilde{A}_{\bm{\alpha}} \mathcal{L}^{\beta}_{\bm{\alpha}} \\
    &\begin{aligned}
        &=\frac{\pi}{4 \sqrt{2}} \Gamma\left(\frac{D-1}{2}\right)\bigg[ 2 D {}_1 F_{1}^{\mathrm{R}} \left(\frac{1}{2}, \frac{D}{2} + 1, \frac{\tilde{d}^2_{\bm{\alpha}}}{2}\right) + \\
        &~~~~\tilde{d}^2_{\bm{\alpha}} {}_1 F_{1}^{\mathrm{R}} \left(\frac{3}{2}, \frac{D}{2} + 2, \frac{\tilde{d}^2_{\bm{\alpha}}}{2}\right)\bigg]
    \end{aligned} \nonumber \\
    &= \begin{cases}
    \displaystyle \frac{\pi^{\frac{3}{2}}}{\sqrt{2}} \e^{\frac{\tilde{d}_{\bm{\alpha}}^2}{4}} I_{0} \left(\frac{\tilde{d}^2_{\bm{\alpha}}}{4}\right) & (D=2), \\
    \displaystyle \frac{\pi^{\frac{3}{2}}}{2 \sqrt{2}} \e^{\frac{\tilde{d}_{\bm{\alpha}}^2}{4}} \left[I_{0} \left(\frac{\tilde{d}^2_{\bm{\alpha}}}{4}\right) - I_{1} \left(\frac{\tilde{d}^2_{\bm{\alpha}}}{4}\right)\right] & (D=4),
    \end{cases} \label{eq:L_A_beta}
\end{align}
where ${}_p F_{q}^{\mathrm{R}}(z)$ is the regularized confluent hypergeometric function.

By performing analytic marginalization, we avoid the need to explore the $KD$-dimensional coefficient space with a numerical method such as \mcmc methods.
This approach significantly reduces the computational cost of the analysis, especially when a large number of \qnms are considered.
Moreover, it naturally accounts for the entire range of coefficient values from $-\infty$ to $\infty$, which is not easily achieved with standard numerical techniques.

\subsection{Sampling procedure} \label{subsec:sampling procedure}

The sampling procedure of our semianalytic method proceeds as follows.
With the analytic marginalization described above, the marginal posterior for the remnant \bh’s mass and spin can be computed as
\begin{equation}
    p(M_f, \chi_f) \propto \pi(M_f, \chi_f) \prod_{\bm{\alpha}} \mathcal{L}^{A,\beta}_{\bm{\alpha}}.
\end{equation}
Samples of $(M_f,\chi_f)$ can then be generated numerically from this posterior distribution.

For each sampled pair $(M_f, \chi_f)$, the mode amplitude $\tilde{A}_{\bm{\alpha}_k}$ is drawn from the conditional distribution,
\begin{equation}
    p(\tilde{A}_{\bm{\alpha}_k}|M_f, \chi_f) = \frac{p(M_f, \chi_f, \tilde{A}_{\bm{\alpha}_k})}{p(M_f, \chi_f)} \propto \mathcal{L}^{\beta}_{\bm{\alpha}_k}.
\end{equation}
Finally, for each sample of $(M_f, \chi_f, \tilde{A}_{\bm{\alpha}_k})$, the angular parameters of the $(D-1)$-dimensional unit sphere are sampled from the likelihood $\mathcal{L}_{\bm{\alpha}_k}$, and the mode coefficients $\{c_{j,\bm{\alpha}_k}\}$ are reconstructed accordingly.

\subsection{Mode identification}

One of the goals of our semianalytic method is to efficiently identify subdominant \qnms associated with $\bm{\alpha}_k$ for $k>0$.
As described in Sec. \ref{sec:gs}, the orthonormal vectors corresponding to $\bm{\alpha}_k$ isolate distinct signal features introduced by the $\bm{\alpha}_k$ modes.
We infer the presence of these modes in the data through their nonzero projection coefficients.
More formally, we consider a mode detected when the marginal posterior $p(\tilde{A}_{\bm{\alpha}_k})$ excludes $\tilde{A}_{\bm{\alpha}_k} = 0$ with a specified credibility level (e.g. $90\%$).
While credible intervals can be constructed in various ways, we adopt the \hpd interval, defined as the smallest continuous interval containing the desired posterior mass.

A similar approach to identify subdominant modes was proposed in \cite{Isi:2021iql}.
Their method involves performing multiple analyses, incrementally adding new modes, and stopping once the added modes are not identified with a specified credibility level.
In contrast, since our method reduces correlations between modes, we infer the presence of the modes with a single analysis including all the modes we are interested in, or at least with fewer analyses.

Another common strategy is to use the Bayes factor to compare models with different numbers of modes.
As noted by Isi et al., this approach is meaningful only when the prior accurately reflects our beliefs about the parameters.
However, because we lack strong prior knowledge about the mode coefficients and allow them to take arbitrary real values, we choose not to use this method.

\section{Application to damped sinusoids} \label{sec:damped}

\begin{figure*}[!ht]
    \centering
    \begin{minipage}{0.48\textwidth}
        \includegraphics[width=\linewidth]{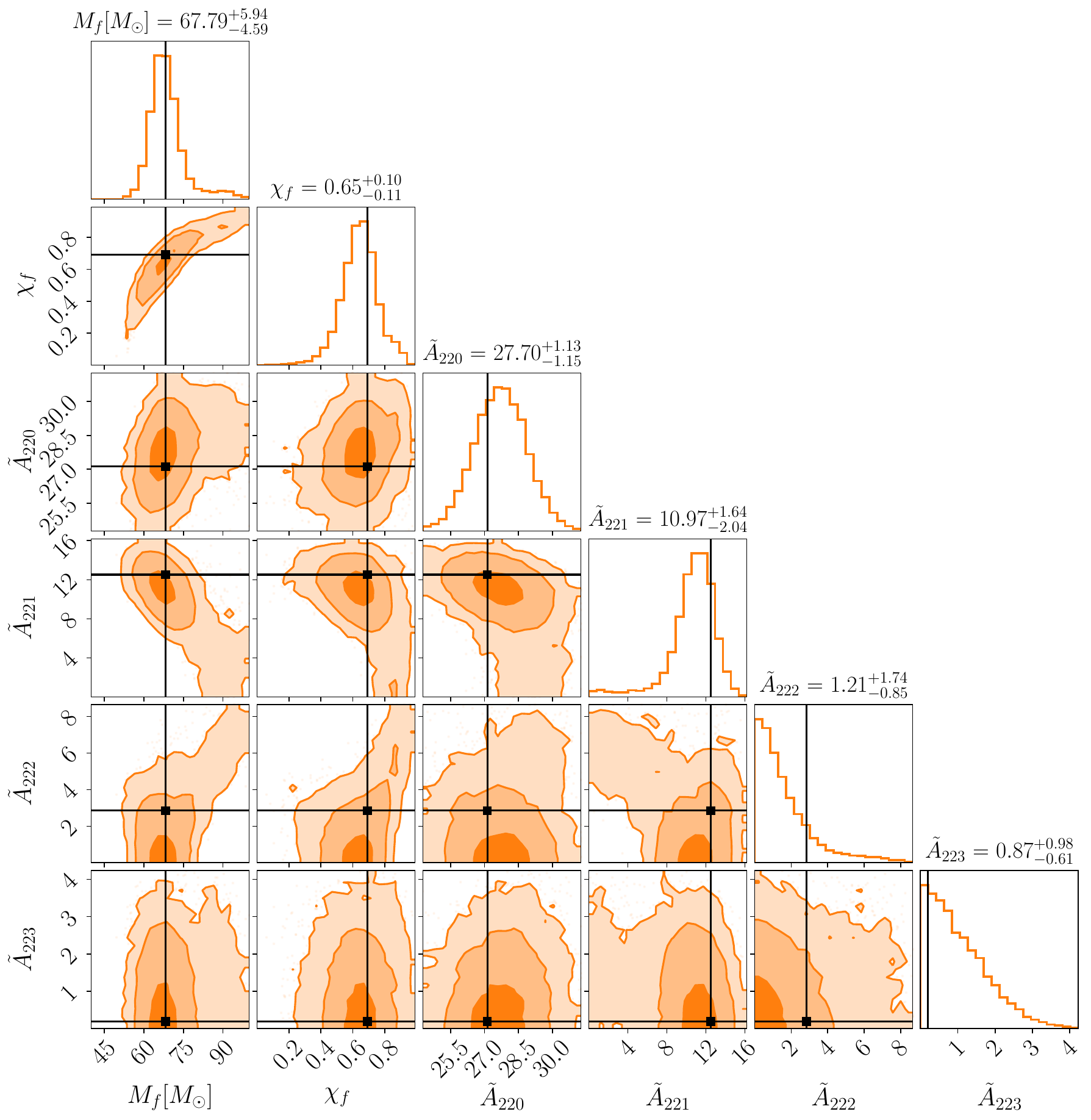}
    \end{minipage}
    \begin{minipage}{0.48\textwidth}
        \includegraphics[width=\linewidth]{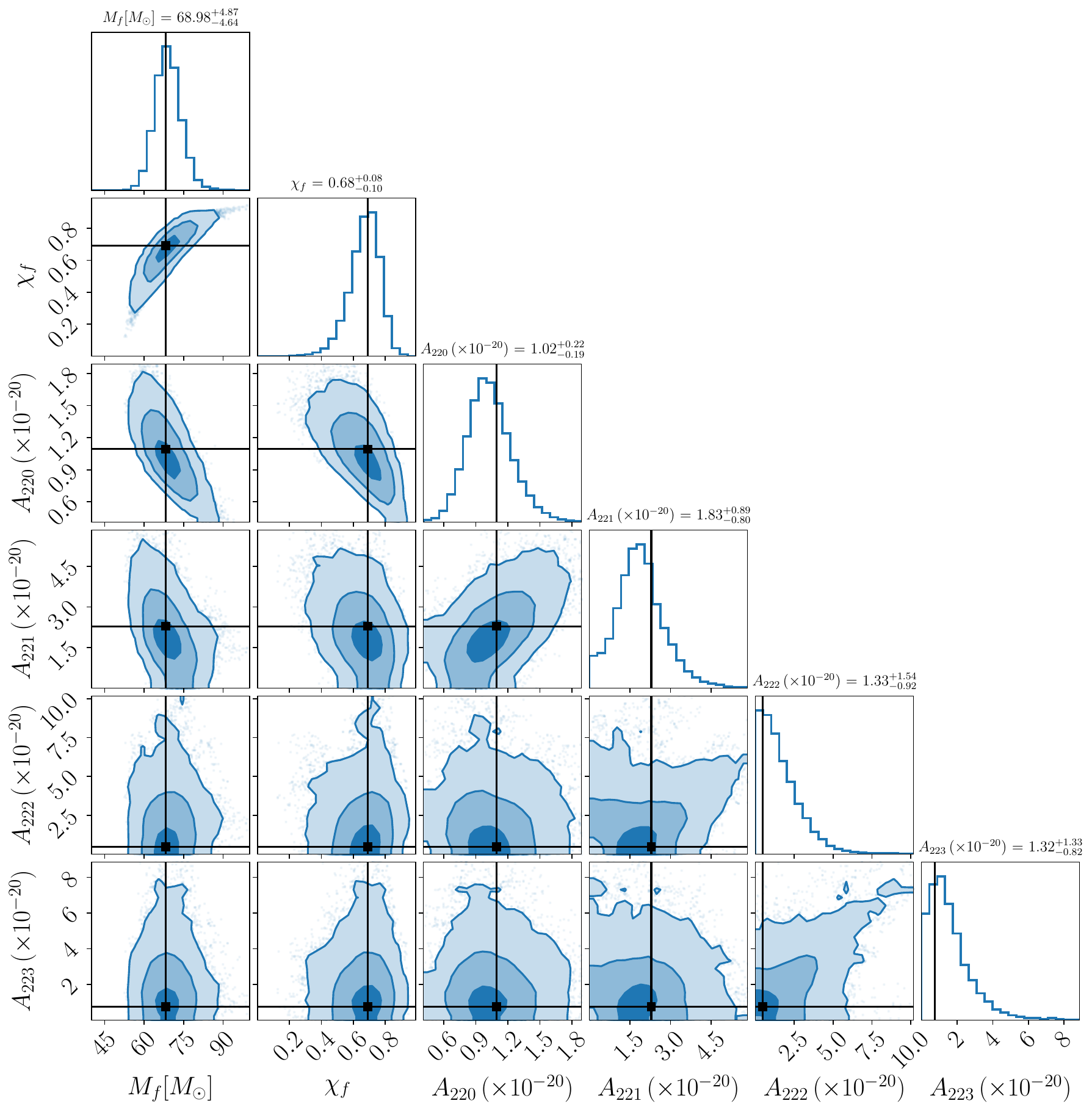}
    \end{minipage}
    \caption{
        Posterior distributions for the remnant \bh's mass, spin, and mode amplitudes obtained by analyzing the four-mode signal.
        The left (right) panel shows the results from the semianalytic (\mcmc) method.
        The contours are color coded to represent the 1-$\sigma$ (39.3\%), 2-$\sigma$ (86.5\%), and 3-$\sigma$ (98.9\%) credible regions.
        The black lines represent the injected values.
    }
    \label{fig:corners_N=3_SNR=30}
\end{figure*}
\begin{figure*}[!ht]
    \centering
    \begin{minipage}{0.48\textwidth}
        \includegraphics[width=\linewidth]{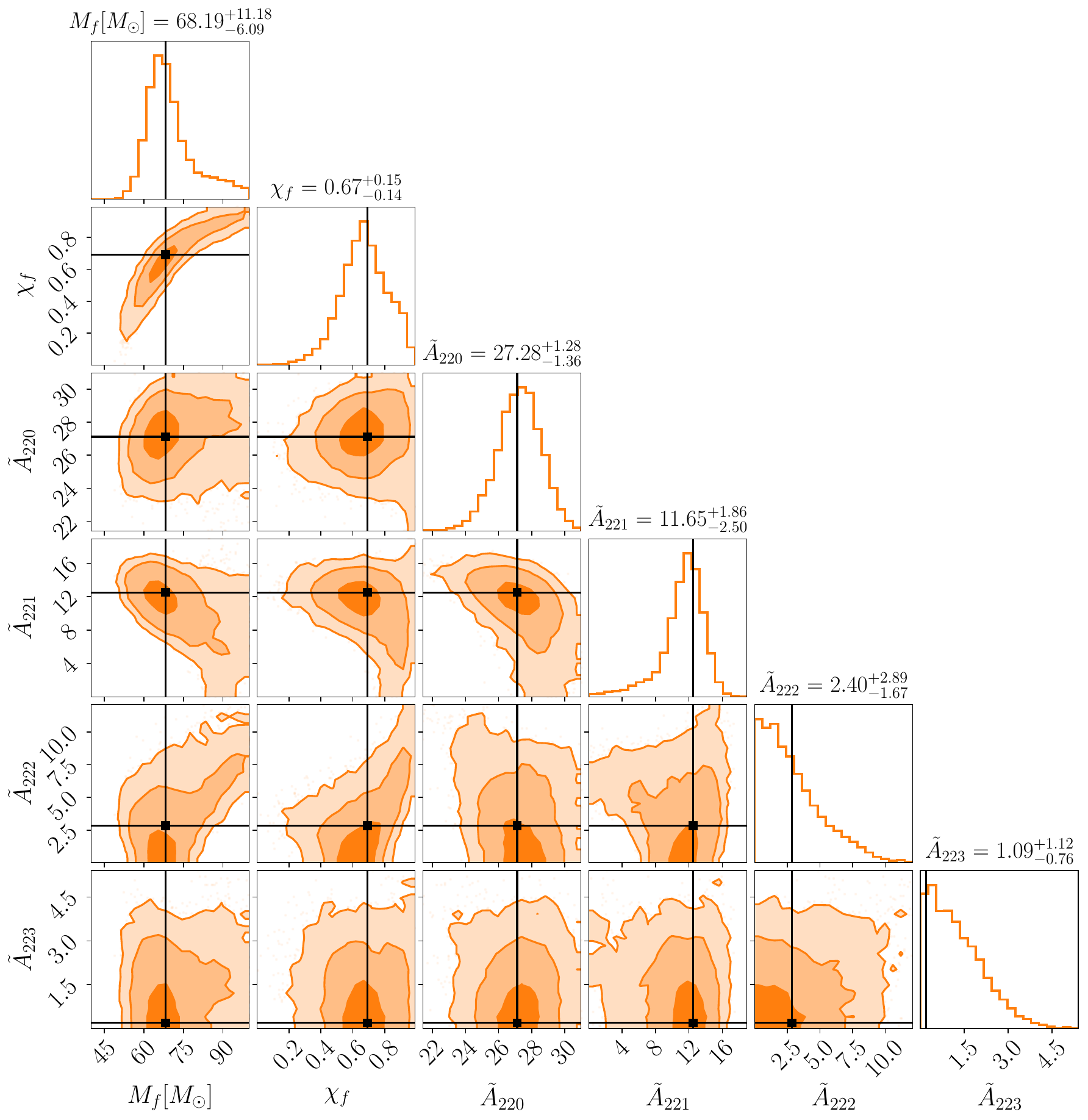}
    \end{minipage}
    \begin{minipage}{0.48\textwidth}
        \includegraphics[width=\linewidth]{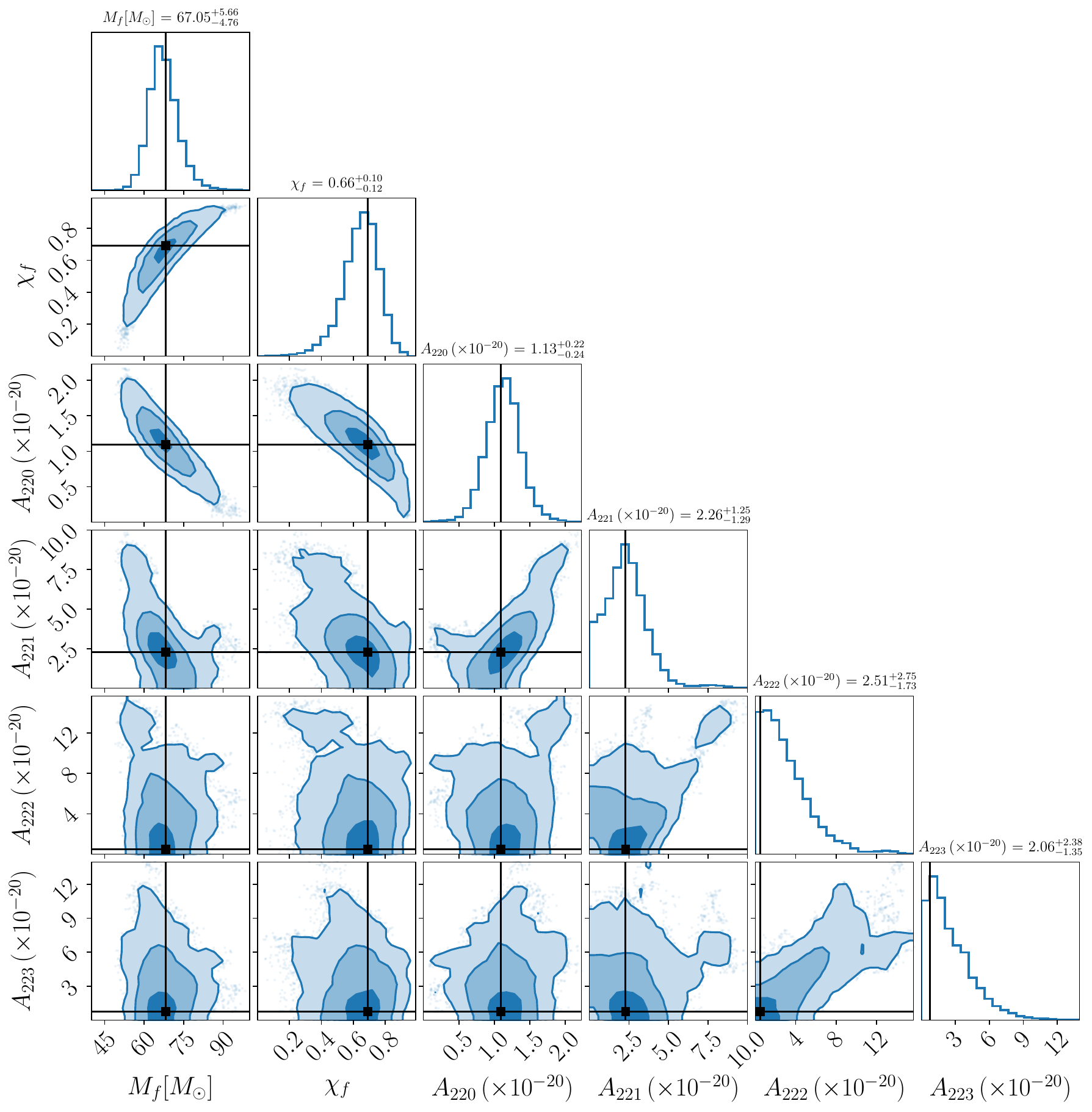}
    \end{minipage}
    \caption{
        Posterior distributions for mass, spin, and mode amplitudes obtained by analyzing the four-mode signal under the assumption of $D=2$ damped sinusoids per \qnm.
        The left (right) panel shows the results from the semianalytic (\mcmc) method.
    }
    \label{fig:corners_N=3_D=2_SNR=30}
\end{figure*}
To evaluate the performance of the analysis using orthonormal modes, we apply it to mock data generated by summing damped sinusoids, as described by Eq.~\eqref{eq:hI3}.
The frequencies and decay times of the damped sinusoids are set to those of Kerr \qnms with the \bh parameters of $M_f=68.2\,M_{\odot}$ and $\chi_f=0.692$, consistent with those inferred from GW150914.
Precise calculations of complex \qnm frequencies at discrete spin values are provided in \cite{Motohashi:2024fwt, motohashi2024kerr}. 
We interpolate their results to compute \qnm frequencies and decay times for both signal injection and parameter inference.

We simulate damped-sinusoidal signals observed by the two LIGO detectors at Hanford and Livingston.
The simulations are performed under the ``zero-noise'' assumption, where the simulated signals are assumed to be observed in the absence of random detector noise.
The sampling rate of data is $2048\,\mathrm{Hz}$, and the duration of analyzed data is $300M$, where $M=68.2\,M_{\odot}$.
The covariance matrix is constructed using the noise \psd estimated with the data around GW150914, available from \cite{LIGO-P1900011-v1}.
The right ascension, declination, and polarization angle are set to values consistent with those of GW150914: $\alpha=1.95\,\mathrm{rad}$, $\delta=-1.27\,\mathrm{rad}$, and $\psi=0.82\,\mathrm{rad}$.
The start time of the damped sinusoids measured at the Earth's center, which we refer to as the peak time $t_{\mathrm{peak}}$, is also chosen to match GW150914: $t_{\mathrm{peak}}=1126259462.423\,\mathrm{s}$, expressed in GPS time.

The simulated signals include only $l=m=2$ modes.
We consider two types of signals: one including four modes up to $n=3$, and another including eight modes up to $n=7$.
These are referred to as the four-mode and eight-mode signals, respectively.
The mode coefficients assumed in these simulations are listed in Tables.~\ref{tab:coefficients_K=4} and \ref{tab:coefficients_K=8}.
For the four-mode signal, the coefficients are chosen to mimic the postmerger phase of SXS:BBH:0305.
For the eight-mode signal, the coefficients are computed based on the results presented in Table I of \cite{Takahashi:2023tkb}, which were obtained by fitting \qnms to a numerical waveform from the \sxs catalog.
Those coefficients are normalized to achieve a network \snr of 30.
Each simulated signal is analyzed assuming the same set of \qnms that is used to construct the corresponding simulated signal.
Unless otherwise specified, the analysis assumes $D=4$ damped sinusoids per \qnm.
All analyses are performed starting from the peak time of the signal.
The true values of right ascension and declination are used to compute the peak time at each detector and the detector beam patterns.

\begin{table}[h]
    \centering
    \caption{
        Injected values of the mode coefficients $c_{j,\alpha_k}$ and amplitude $A_{\alpha_k}$ for the four-mode signal.
        The listed values are scaled by $10^{-20}$.
        The coefficients are chosen to mimic the postmerger phase of SXS:BBH:0305 and to achieve a network \snr of 30.
    }
    \setlength{\tabcolsep}{7pt}
    \begin{tabular*}{\linewidth}{@{\hspace{5pt}}@{\extracolsep{\fill}}cccccc}
        \hline\hline
        $k$ & $c_{0,\alpha_k}$ & $c_{1,\alpha_k}$ & $c_{2,\alpha_k}$ & $c_{3,\alpha_k}$ & $A_{\alpha_k}$ \\
        \hline
        \rule{0pt}{2.5ex}%
        0 & $-0.4768$ & $0.8502$ & $-0.3376$ & $-0.3609$ & $1.0929$ \\
        1 & $2.0116$ & $-0.7439$ & $-0.2217$ & $0.7646$ & $2.2877$ \\
        2 & $-0.4013$ & $-0.0456$ & $0.2399$ & $-0.0094$ & $0.46989$ \\
        3 & $-0.7062$ & $-0.1697$ & $0.1110$ & $-0.0951$ & $0.7409$ \\
        \hline\hline
    \end{tabular*}
    \label{tab:coefficients_K=4}
\end{table}
\begin{table}[h]
    \centering
    \caption{
        Injected values of the coefficients $c_{j,\bm{\alpha}}$ and amplitudes $A_{\alpha_k}$ for the eight-mode signal.
        The listed values are scaled by $10^{-20}$.
        The coefficients are chosen to mimic the postmerger phase of SXS:BBH:0305 and to achieve a network SNR of 30.
        The values of $c_{2,\alpha_k}$ and $c_{3,\alpha_k}$ are determined from the relations $c_{2,\alpha_k}=c_{1,\alpha_k}$ and $c_{3,\alpha_k}=-c_{0,\alpha_k}$, which correspond to the circular polarization condition.
    }
    \setlength{\tabcolsep}{7pt}
    \begin{tabular*}{0.8\linewidth}{@{\hspace{5pt}}@{\extracolsep{\fill}}cccc}
        \hline\hline
        $k$ & $c_{0,\alpha_k}$ & $c_{1,\alpha_k}$ & $A_{\alpha_k}$ \\
        \hline
        \rule{0pt}{2.5ex}%
        0 & $0.1155$ & $-1.2096$ & $1.7184$ \\
        1 & $4.1100$ & $3.2590$ & $7.4180$ \\
        2 & $-13.5496$ & $-3.7801$ & $19.8937$ \\
        3 & $27.3075$ & $7.6071$ & $40.0891$ \\
        4 & $-35.6762$ & $-19.3948$ & $57.4274$ \\
        5 & $25.5084$ & $25.8310$ & $51.3403$ \\
        6 & $-8.4943$ & $-15.5784$ & $25.0934$ \\
        7 & $1.1162$ & $3.4913$ & $5.1837$ \\
        \hline\hline
    \end{tabular*}
    \label{tab:coefficients_K=8}
\end{table}
For comparison, we also perform an analysis with flat priors on the amplitude of the original (nonorthogonal) mode $A_{\bm{\alpha}_k}$ defined in Eq.~\eqref{eq:A}.
In this analysis, the amplitude prior is taken to be uniform over a finite range, $(0,2.5\times10^{-19})$, and we use the Metropolis-Hastings sampler as implemented in the Python package \textsc{PyMC} \cite{17cb5e2c4e294d189fb3c54fbdc31562} to draw samples from the posterior distribution.
In both analyses, we adopt uniform priors for mass and spin as $M_f/M_{\odot}\sim\mathcal{U}(40,100)$ and $\chi_f\sim\mathcal{U}(0,0.99)$, respectively.

\subsection{Results} \label{sec:result}

\begin{figure*}
    \centering
    \includegraphics[width=\linewidth]{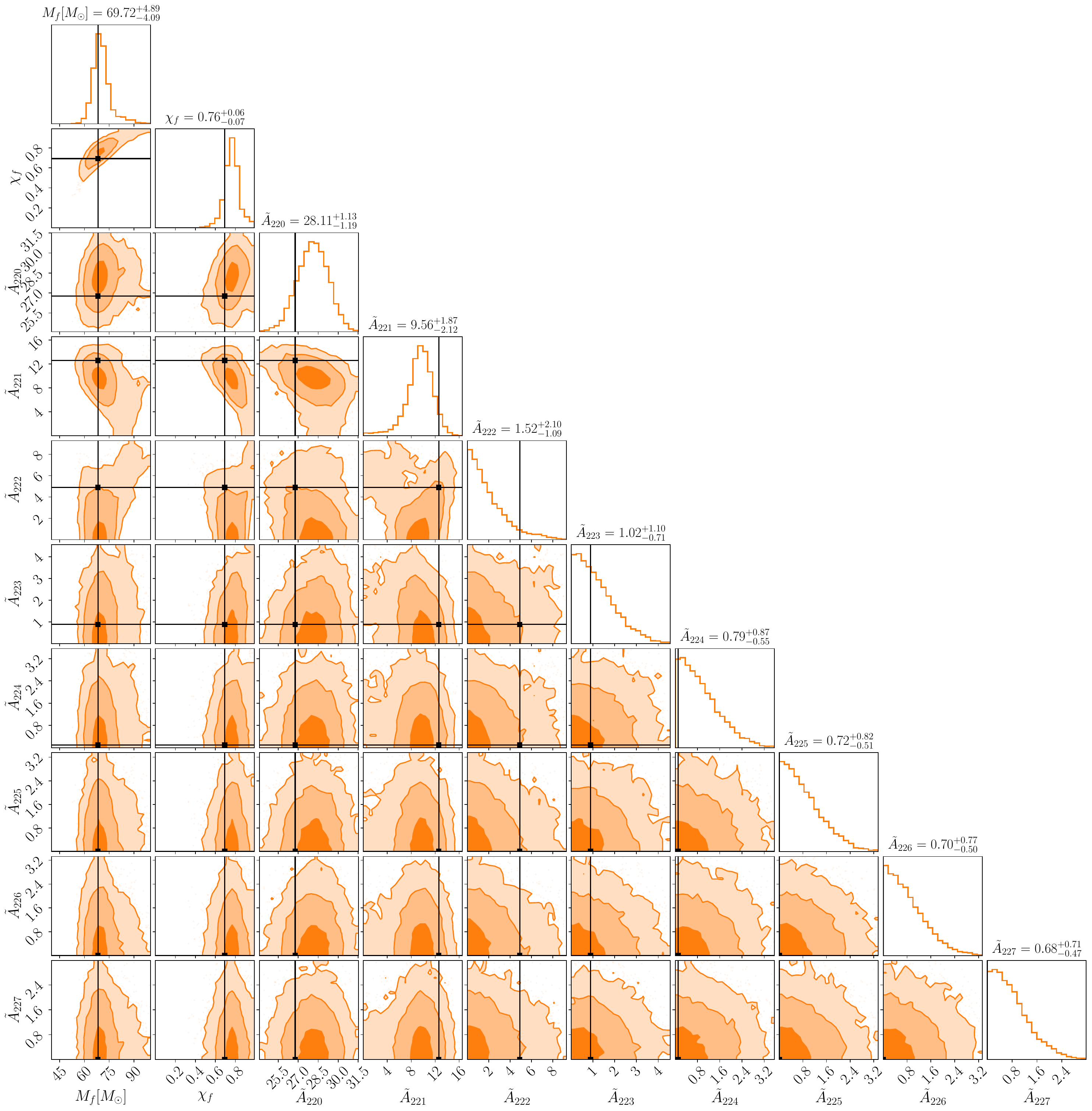}
    \caption{
        Posterior distributions for mass, spin, and mode amplitudes obtained from the semianalytic method by analyzing the eight-mode signal.
        The analysis starts from the peak time.
    }
    \label{fig:N=7_SNR=30}
\end{figure*}
Figure \ref{fig:corners_N=3_SNR=30} shows the posterior distributions of the remnant \bh's mass, spin, and mode amplitudes obtained by analyzing the four-mode signal.
The left panel presents the results from our semianalytic method, while the right panel shows the reference \mcmc analysis. 
As seen in the right panel, $A_{221}$ and $A_{222}$ are anticorrelated, and the probability of both vanishing simultaneously is small.
More quantitatively, the hypothesis $A_{221}=A_{222}=0$ is excluded at the 99\% credible level.
However, the fact that one of them is likely not zero is not apparent in either of their one-dimensional distributions, as neither excludes $A_{221}=0$ or $A_{222}=0$.
On the other hand, in the posterior distributions obtained from our semianalytic method, the one-dimensional posterior distribution of $\tilde{A}_{221}$ excludes $\tilde{A}_{221}=0$ from the 97.6\% \hpd credible region, indicating that the $(2,2,1)$ mode can be extracted from the data based solely on the one-dimensional marginal distribution.
\begin{figure*}[!ht]
    \centering
    \begin{minipage}{0.43\textwidth}
        \includegraphics[width=\linewidth]{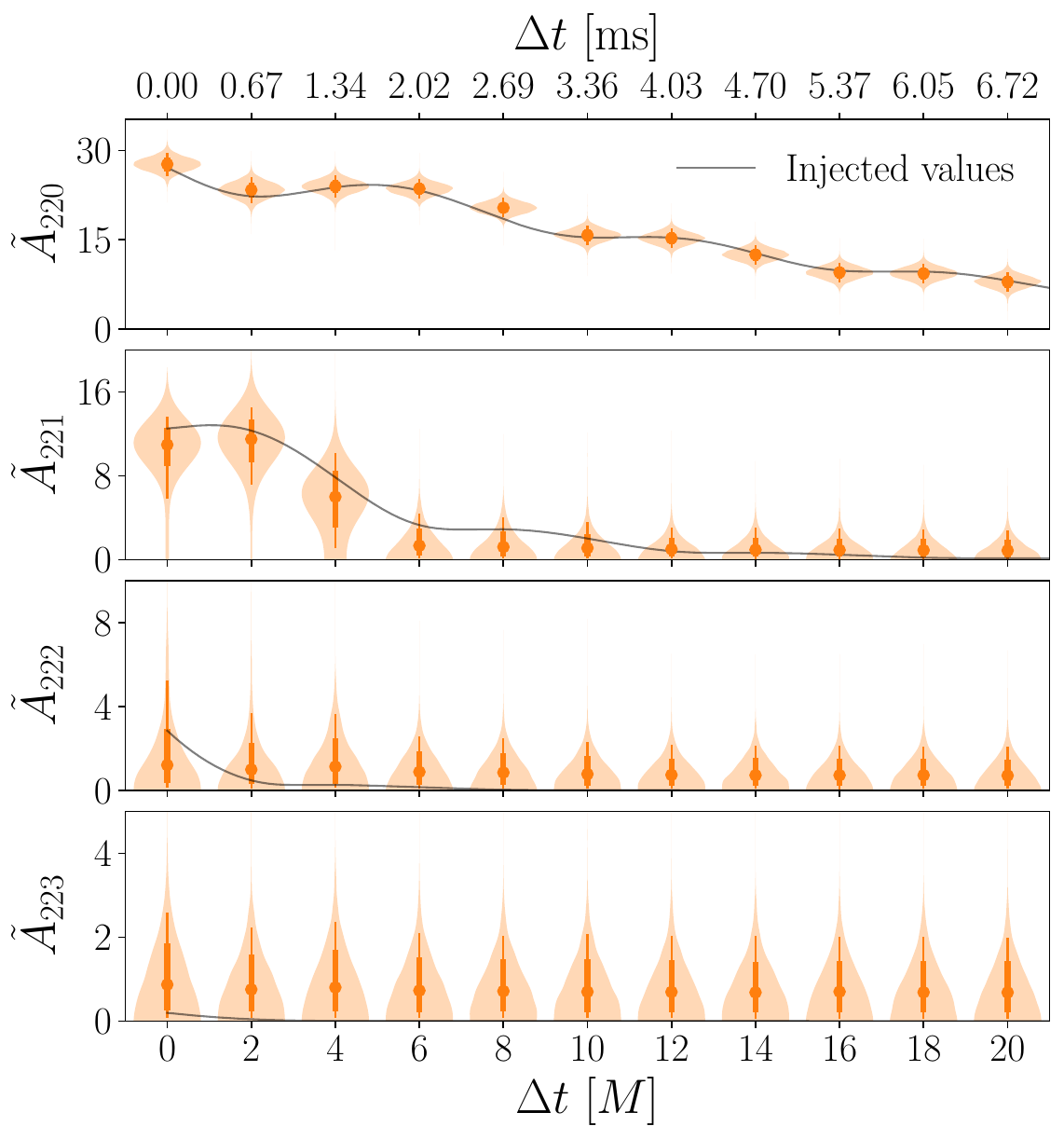}
    \end{minipage}
    \begin{minipage}{0.43\textwidth}
        \includegraphics[width=\linewidth]{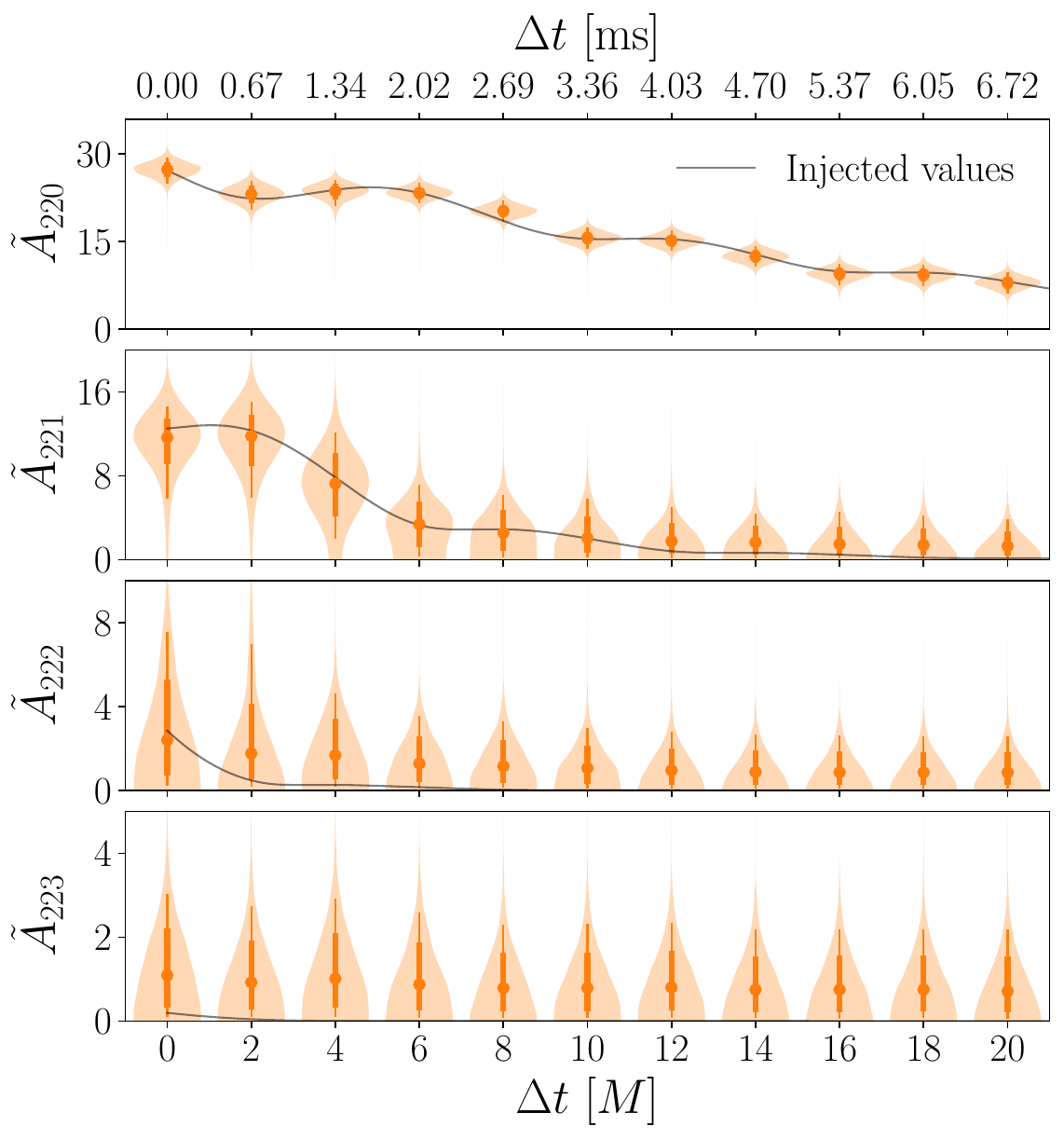}
    \end{minipage}
    \caption{
        Posterior distributions for the mode amplitudes obtained from the semianalytic method by analyzing the four-mode signal, evaluated using data segments starting at different times.
        The left (right) panel shows the result of the analysis assuming $D=4$ ($D=2$) damped sinusoids per \qnm.
        The horizontal axes indicate the start time relative to the signal peak.
        Circles, thick vertical lines, and thin vertical lines denote the median, 1-$\sigma$ (68\%), and 2-$\sigma$ (95\%) credible intervals, respectively.
        The black solid lines represent the injected values.
        The oscillatory behavior of the true values, rather than a simple exponential decay, arises because the basis vectors are linear combinations of different \qnms.
    }
    \label{fig:violins_K=4}
\end{figure*}

Figure~\ref{fig:corners_N=3_D=2_SNR=30} shows the result of analyzing the four-mode signal assuming $D=2$ damped sinusoids per \qnm.
The posterior distributions obtained assuming $D=2$ qualitatively resemble those obtained assuming $D=4$.
Because the two LIGO detectors are nearly co-aligned, as discussed in Sec.~\ref{sec:review}, the signal model with $D=2$ can describe the observed signal with sufficient accuracy.
In the $D=2$ MCMC analysis, we observe strong positive correlations between amplitude parameters, such as between $A_{222}$ and $A_{223}$.
These modes tend to be out of phase so that they cancel each other out in the resulting waveform.
As seen in the results for $D=4$, the correlations between the amplitude parameters are mitigated with our semianalytic method, and the one-dimensional posterior distribution of $\tilde{A}_{221}$ excludes $\tilde{A}_{221}=0$ at the 99.7\% credibility.

As the number of modes in the template increases, correlations between mode amplitudes become more severe.
This tends to lead to strong degeneracies between the different mode amplitudes, making it difficult to extract individual mode amplitudes in a robust way~\cite{Clarke_2024}.
Figure \ref{fig:N=7_SNR=30} shows the results obtained by applying our semianalytic method to the eight-mode signal.
Even with a large number of modes, the amplitude parameters of the orthogonal modes remain only weakly correlated.
The one-dimensional posterior distribution of $\tilde{A}_{221}$ excludes $\tilde{A}_{221}=0$ at the 99.5\% credibility.

We also observe that the peak of the marginal posterior distribution for $M_f$ and $\chi_f$ is off from their true values in Fig.~\ref{fig:N=7_SNR=30}.
Although the likelihood function is maximized at the true parameter values in the zero-noise simulation, the marginal posterior does not necessarily peak at those values.
In our case, the product of the terms involving the Bessel functions in Eq.~\eqref{eq:L_A_beta} taken over different $\bm{\alpha}_k$ does not necessarily get maximized at the true values of $M_f$ and $\chi_f$.
This accounts for the observed shift in the marginal posterior for $M_f$ and $\chi_f$.

Similarly, we observe that the peak of the marginal posterior distribution for $\tilde{A}_{\bm{\alpha}_k}$ is off from their true values in Figs.~\ref{fig:corners_N=3_SNR=30}, \ref{fig:corners_N=3_D=2_SNR=30}, and \ref{fig:N=7_SNR=30}.
This offset is likely attributed to the marginal likelihood over $\tilde{A}_{\bm{\alpha}_k}$ in Eq.~\eqref{eq:L_beta}.
Fortunately, this offset predominantly shifts the posterior distribution toward values smaller than the true ones, and therefore rarely leads to critical false-positive results.
A detailed discussion of these offsets is provided in the Appendix \ref{append:offsets}.

Figure \ref{fig:violins_K=4} shows the posterior distributions of the mode amplitudes obtained by applying the semianalytic method to a four-mode signal while varying the analysis start time $\Delta t$ from the signal peak.
For all choices of the analysis start time, and in particular for late times when the template waveform contains more modes than are actually present in the signal, we observe no indication of false positives in the inferred mode amplitudes.
For completeness, as shown in Appendix \ref{append:additional tests}, we also perform additional tests in which the eight-mode signal is analyzed using a four-mode template, and conversely, the four-mode signal is analyzed using an eight-mode template.
The former setup reflects a more realistic situation, where the template contains fewer modes than are actually present in the signal.
As shown there, neither of these tests likewise leads to any false positives.

\subsection{Reweighting for different prior choices}

In principle, we can obtain the samples under a flat prior on $\{A_{\bm{\alpha}_k}\}$ by reweighting samples generated under a flat prior on $\{\tilde{A}_{\bm{\alpha}_k}\}$ employing the rejection sampling or importance sampling techniques.
The sampling procedure proceeds as follows: First, generate a sample set $\{M_f,\chi_f,\{\tilde{c}_{i,\bm{\alpha}_k}\}\}$ using our semianalytic method.
Then, convert $\{\tilde{c}_{j,\bm{\alpha}_k}\}$ into $\{c_{j,\bm{\alpha}_k}\}$ using the inverse of Eq.~\eqref{eq:amplitude_transformation}, evaluated at the corresponding $\{M_f,\chi_f\}$.
Finally, perform rejection or importance sampling by assigning the following weight to each sample:
\begin{equation}
    w = \left|\det U\right| \prod_{0\leq k\leq K-1} \qty(\frac{\tilde{A}_{\bm{\alpha}_k}}{A_{\bm{\alpha}_k}})^{D-1}.
\end{equation}

\begin{figure}[h!]
    \centering
    \includegraphics[width=\linewidth]{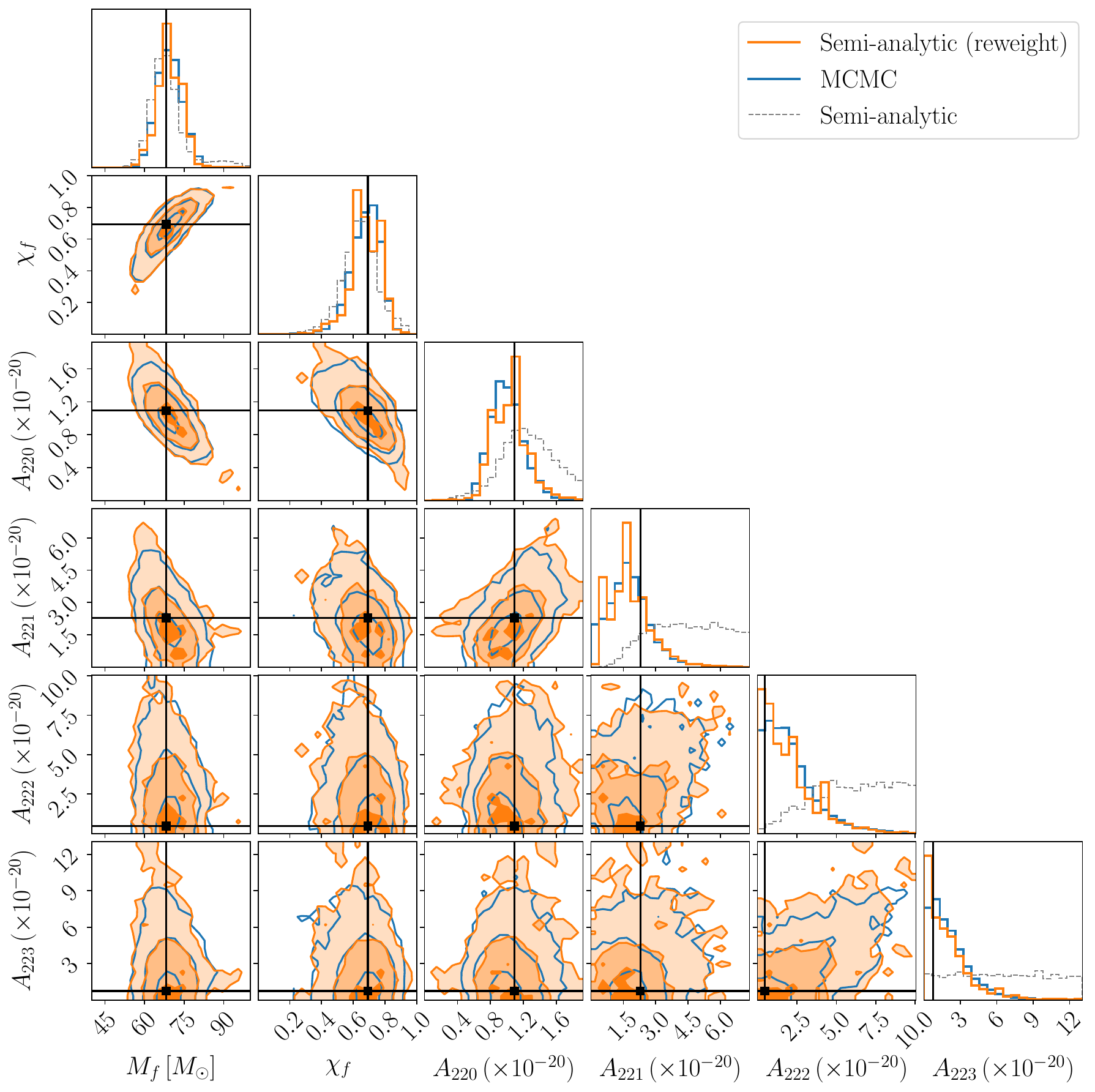}
    \caption{
        Comparison of posterior distributions obtained by reweighting samples from our semianalytic method (orange filled contours) and those from the reference \mcmc method (blue lines).
        The gray dashed lines represent one-dimensional posteriors of the original samples from our semianalytic method.
        The results are obtained from the analysis of the four-mode signal assuming $D=4$ damped sinusoids per \qnm.
        The contours represent the 1-$\sigma$ (39.3\%), 2-$\sigma$ (86.5\%), and 3-$\sigma$ (98.9\%) credible regions.
        The black lines represent the injected values.
    }
    \label{fig:reweight}
\end{figure}

However, the weight $w$ diverges when any of the mode amplitudes $A_{\bm{\alpha}_k}$ vanish.
Furthermore, as the number of \qnms increases, $w$ grows more rapidly for small $A_{\bm{\alpha}_k}$, resulting in an unstable distribution with large variance.
Figure \ref{fig:reweight} shows the results of applying importance sampling to the four-mode signal assuming $D=4$, in comparison with the results from the reference \mcmc analysis.
The importance sampling is performed using $2\times10^6$ samples.
However, due to the reason discussed above, the reweighted distributions tend to be less smooth or exhibit different structures than the \mcmc results, particularly in the region of small $A_{\bm{\alpha}_k}$.
Nevertheless, due to its speed and ease of parallelization, the semianalytic method combined with rejection or importance sampling can be a useful approach, particularly when the number of \qnms considered is small.

\section{Application to binary black hole waveforms} \label{sec:sxs}

\begin{figure*}
    \centering
    \begin{minipage}{0.49\textwidth}
\includegraphics[width=\linewidth]{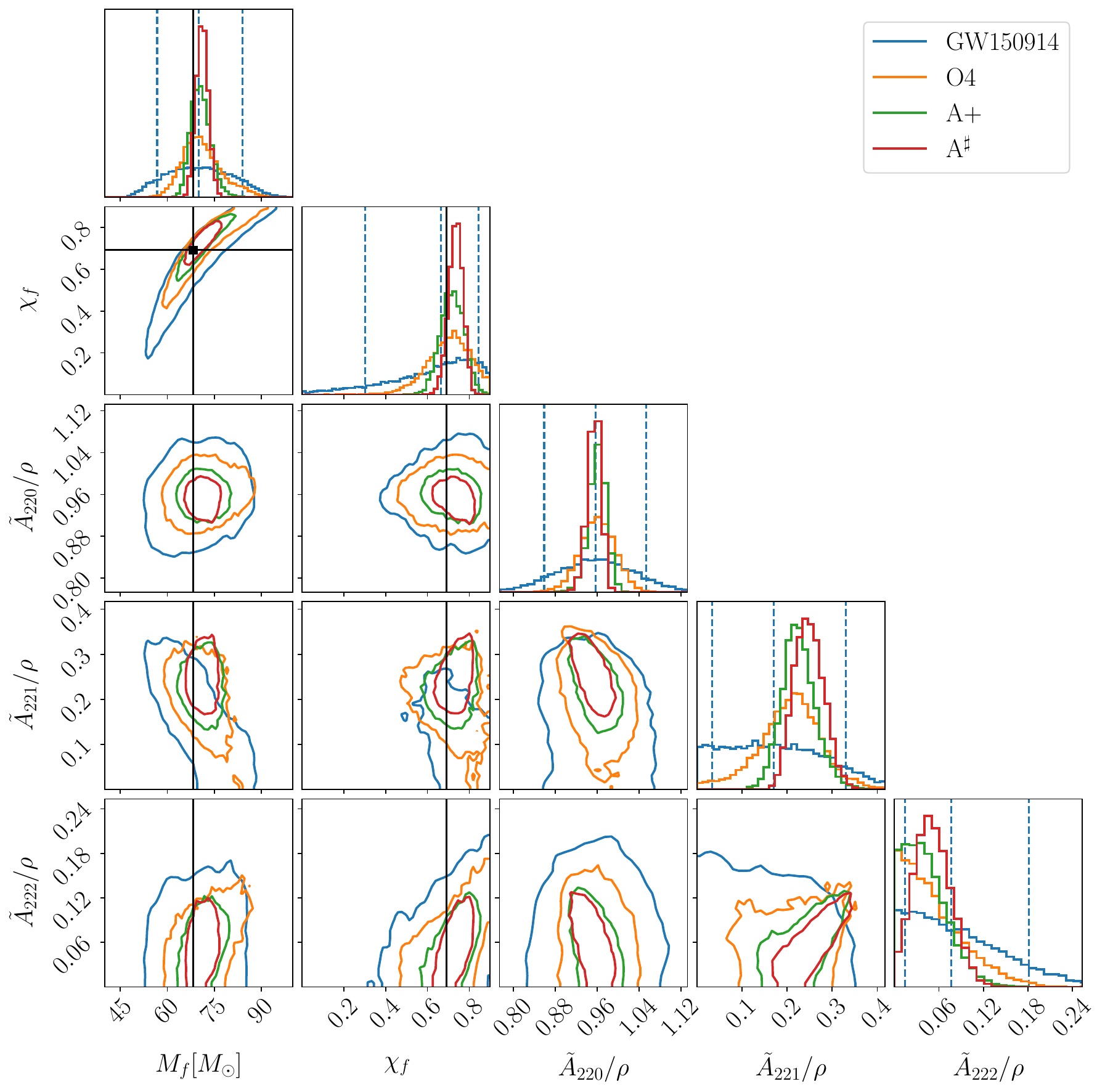}
    \end{minipage}
    \hfill
    \begin{minipage}{0.49\textwidth}
        \includegraphics[width=\linewidth]{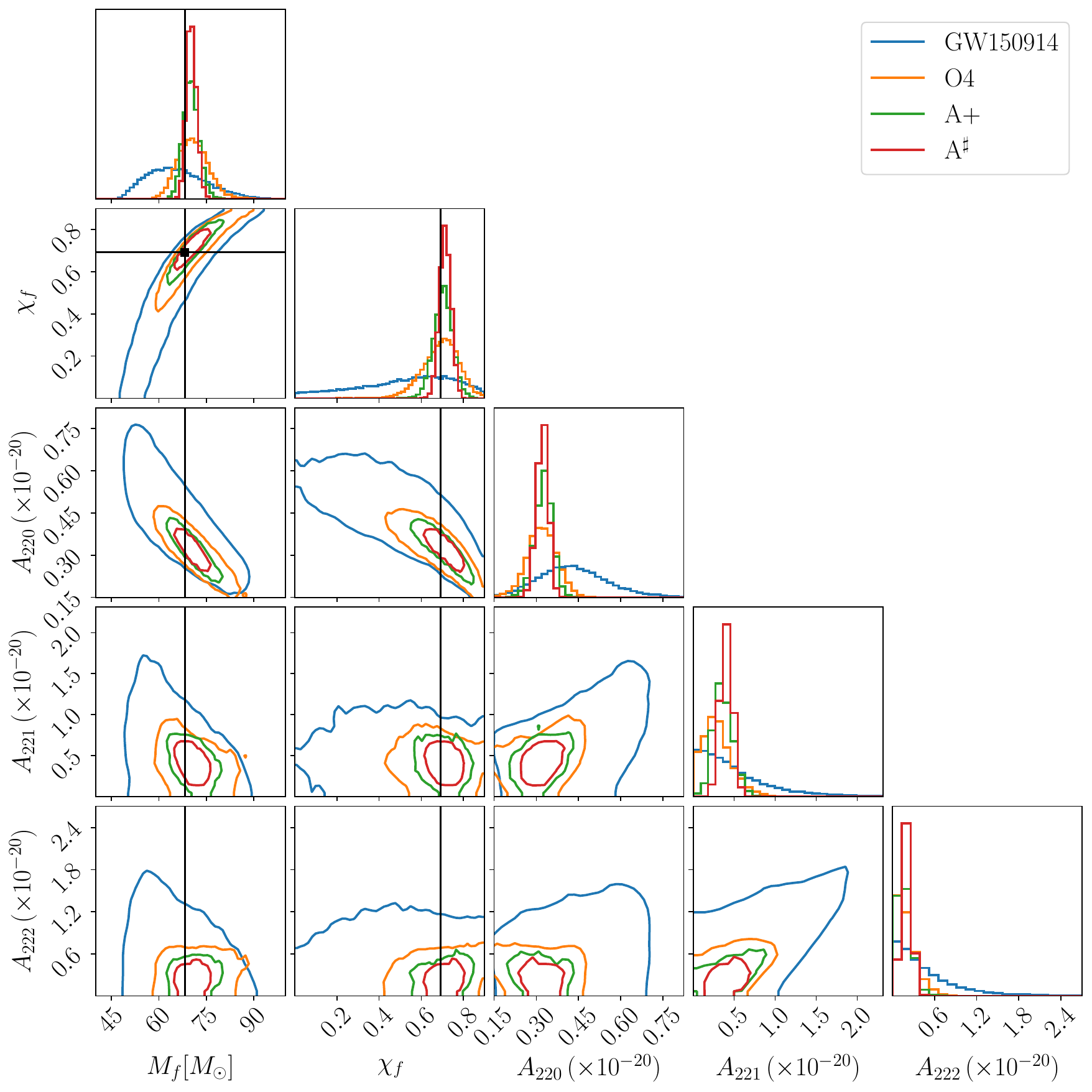}
    \end{minipage}
    \caption{
        Posterior distributions for mass, spin, and mode amplitudes, evaluated using the $l=|m|=2$ spherical harmonic modes from \sxs:\bbh:0305.
        The analysis start time at each detector is set to $5M$ after the peak time, where $M=68.2\,M_{\odot}$.
        The left panel shows the results obtained from our semianalytic method.
        By definition, the quantity $\tilde{A}_{\bm{\alpha}_k}$ is normalized by the detector noise.
        Therefore, the values of the mode amplitude are scaled by dividing by the optimal network \snr $\rho$ computed over the data segment beginning at the analysis start time for each detector sensitivity.
        The right panel shows the results obtained from the reference \mcmc method.
        The blue, orange, green, and red lines present the results assuming the noise \psd estimated using the data around GW150914, the design sensitivity for the O4, the A+ sensitivity, and the $\mathrm{A}^{\sharp}$ sensitivity, respectively.
        The contours represent 90\% credible regions.
    }
    \label{fig:corners_sxs}
\end{figure*}
Next, we apply our semianalytic method to the numerical-relativity waveform provided in the latest \sxs catalog (v3.0.0)~\cite{Boyle_2019,scheel2025sxscollaborationscatalogbinary}.
Specifically, we employ the waveform from the \sxs:\bbh:0305 simulation, which models a nonprecessing \bbh merger with a mass ratio of 1.22 and a dimensionless spin magnitude of 0.692, parameter values consistent with those inferred from GW150914.
We set the other parameters such as detector-frame final mass $M_f$, peak time measured at the Earth's center $t_{\mathrm{peak}}$, luminosity distance $d_{\mathrm{L}}$, right ascension $\alpha$, declination $\delta$, polarization angle $\psi$, inclination angle $\iota$, and constant phase $\phi$ to the values consistent with those inferred from GW150914,
\begin{align}
    &M_f=68.2\,M_{\odot},~~~d_{\mathrm{L}}=390\,\mathrm{Mpc}, \nonumber \\
    &\alpha=1.95\,\mathrm{rad},~~~\delta=-1.27\,\mathrm{rad},~~~\psi=0.82\,\mathrm{rad}, \nonumber \\
    &\iota=0,~~~\phi=0,~~~t_{\mathrm{peak}}=1126259462.423\,\mathrm{s}. \nonumber
\end{align}
The peak time is defined as the time at which $|h_+(t)|^2 + |h_\times(t)|^2$ gets maximized.
The simulated signal only contains the $l=|m|=2$ spherical harmonic modes from the \sxs:\bbh:0305 dataset.

As in the previous Section, we perform zero-noise simulations.
We assume a two-detector network consisting of the LIGO Hanford and Livingston.
We consider four representative sensitivity curves: the noise \psd estimated with the data around GW150914 \cite{LIGO-P1900011-v1}, the design sensitivity for the \ofour~\cite{LIGO-T1800044-v5}, the anticipated design sensitivity for the \ofive (A+ sensitivity)~\cite{LIGO-T1500293-v13}, and the proposed $\mathrm{A}^{\sharp}$ sensitivity for future upgrades~\cite{LIGO-T2300041-v1}.
The sampling rate of the data is 2048\,Hz, and the duration of the analyzed data is $300M$, where $M=68.2\,M_{\odot}$.
In this setup, assuming the sensitivity of each detector, the optimal network \snrs of the postpeak signal---defined by $\rho=\sqrt{(\bm{d},\bm{d})}$---are 14, 34, 74, and 121.

We adopt uniform priors for mass and spin as $M_f/M_{\odot}\sim\mathcal{U}(40,100)$ and $\chi_f\sim\mathcal{U}(0,0.9)$, respectively.
The analysis start time at each detector is set to $5M$ after the peak time, where $M=68.2\,M_{\odot}$.
As in the previous section, we use the complex \qnm frequencies provided in \cite{Motohashi:2024fwt, motohashi2024kerr} to compute \qnm frequencies and decay times.
The analysis takes into account \qnms with $(l,m,n)=(2,2,0),~(2,2,1),~(2,2,2)$ and $D=2$ damped sinusoids per \qnm.
As we do in the previous Section, we perform an \mcmc analysis using flat priors on the original mode amplitudes for comparison.
The amplitude prior is assumed to be uniform over $(0,2.5\times10^{-19})$, and we use the same sampler as in the previous Section to draw samples.

Figure \ref{fig:corners_sxs} shows the posterior distributions for each detector sensitivity, where the left and right panels show the results from our semianalytic method and the reference \mcmc method, respectively.
By construction, $\tilde{A}_{\bm{\alpha}}$ represents the amplitude of the template normalized by the detector noise.
Taking this effect into account, we present the mode amplitudes normalized by the optimal network \snr $\rho$ computed over the segment beginning at the analysis start time for each detector sensitivity.

Comparing the left and right panels of Fig.~\ref{fig:corners_sxs}, we observe that, as confirmed by the analyses of damped sinusoids, the use of orthonormal modes reduces correlations between modes---for example, between $(2, 2, 0)$ and $(2, 2, 1)$.
This allows us to identify subdominant modes from their one-dimensional distributions with high credibility.
For example, the one-dimensional posterior distribution of $A_{221}$ assuming the O4 sensitivity shows that the hypothesis $A_{221}=0$ is disfavored at the 80\% credible level.
On the other hand, when the semianalytic method is used, the reduced correlation leads to the hypothesis $\tilde{A}_{221}=0$ being rejected at the 98\% credible level.
These results demonstrate that our semianalytic method is effective for identifying subdominant \qnms from \bbh signals.

\begin{figure}[h]
    \centering
    \includegraphics[width=\linewidth]{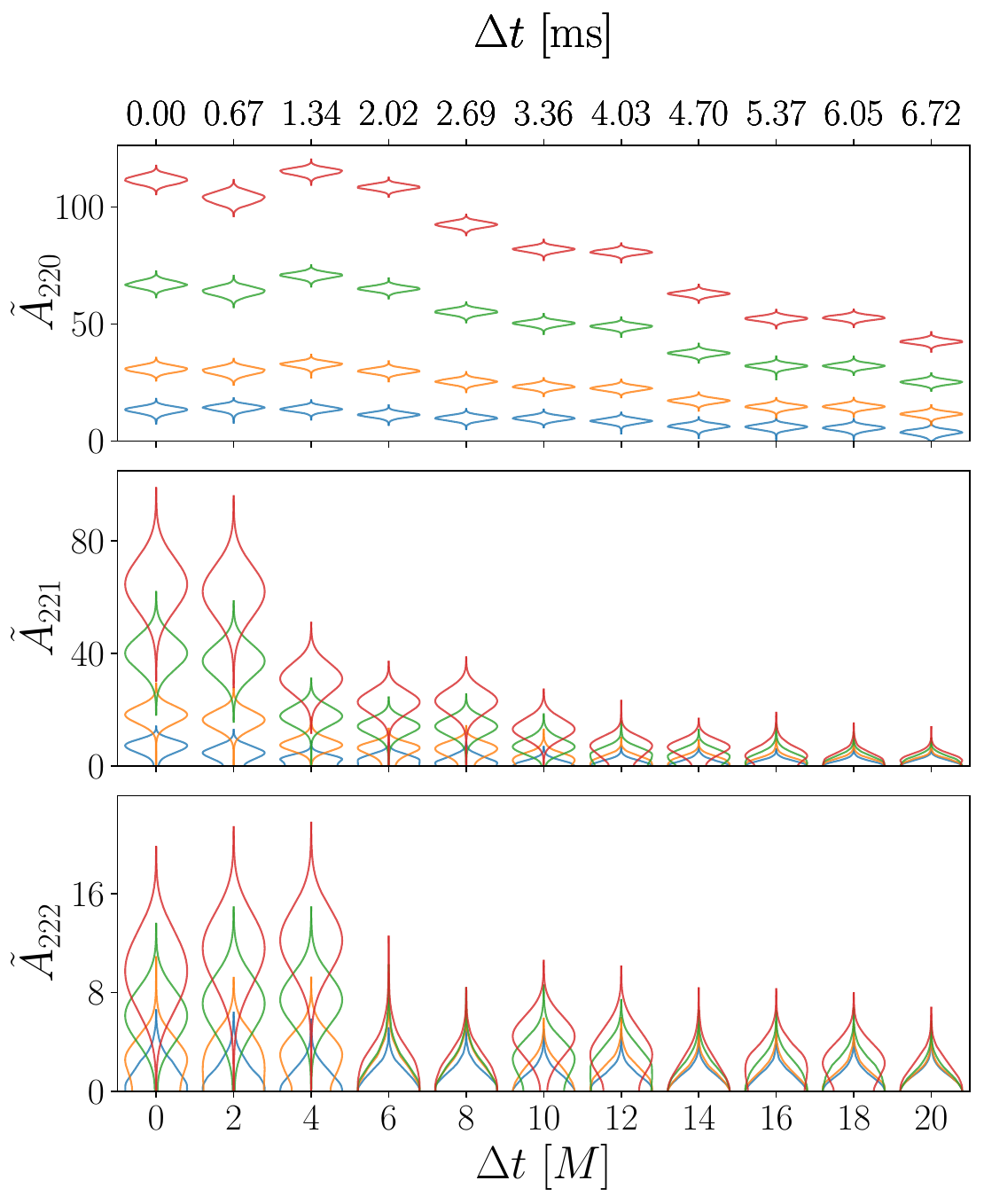}
    \caption{
        Posterior distributions for the mode amplitudes obtained from the semianalytic method by analyzing the $l=|m|=2$ spherical harmonic modes from \sxs:\bbh:0305.
        The horizontal axes show the results obtained with different analysis start times.
        The blue, orange, green, and red lines present the results assuming the noise \psd estimated using the data around GW150914, the design sensitivity for the O4, the A+ sensitivity, and the $\mathrm{A}^{\sharp}$ sensitivity, respectively.
    }
    \label{fig:violin_plot_sxs}
\end{figure}
We then vary the start time of the analysis, defined as the time $\Delta t$ measured from the peak time, within the range $\Delta t\in[0,20]M$.
The corresponding results are presented in Fig.~\ref{fig:violin_plot_sxs}.
The figure shows that overtones can be recovered with larger amplitudes when the analysis begins shortly after the strain peak.
In particular, for relatively high detector sensitivity such as A+ and $\mathrm{A}^{\sharp}$, the presence of $\tilde{A}_{221}$ is indicated even for $\Delta t=10M$, where the linear \qnm analysis is expected to be robust.
Furthermore, for these sensitivities, the hypothesis $\tilde{A}_{222}=0$ is disfavored with high probability for $\Delta t \leq 4M$.
However, we note that very close to the peak, nonlinear effects may also be present.
We also find that the amplitude distributions display an oscillatory pattern with varying start times, consistent with the behavior observed in Fig.~\ref{fig:violins_K=4}.

\section{Conclusion} \label{sec:conclusion}

In this paper, we propose an efficient Bayesian method for identifying multiple \qnms in \gws from \bbh mergers.
Our approach applies the Gram-Schmidt algorithm to the \qnm basis to construct orthonormal modes, thereby reducing correlations among them.
This reduction in correlation is demonstrated through numerical experiments presented in Secs. \ref{sec:damped} and \ref{sec:sxs}.
By adopting a prior that is uniform over the amplitudes of the orthonormal modes, we enable analytical marginalization over the amplitudes, which significantly reduces the computational cost.
Our method is particularly well suited for analyzing future \bbh events with high \snrs, where the detection of multiple \qnms is more likely.

The identification of overtones with less correlation would be especially valuable for high-spin Kerr ringdowns, where the \qnm frequencies of the overtones get close.
It would be interesting to apply our method to this regime and investigate extraction of resonant excitations near exceptional points~\cite{Motohashi:2024fwt}.

While this work focuses on mode identification, once multiple \qnms are observed, the next step is to test \gr by checking the consistency of their frequencies and damping times. Previous studies have addressed this by introducing additional parameters to allow the frequencies and damping times of subdominant modes to deviate from their GR-predicted values \cite{Isi:2019aib,LIGOScientific:2020tif,LIGOScientific:2021sio}.
The extension of our method to accommodate such analyses is a future work.

Another promising direction is the application of our method to real data.
It would be interesting to revisit existing events, such as GW150914, to investigate the significance of subdominant modes with our method.
With future improvements in detector sensitivity, \bbh events with higher \snrs will become more common, making our method increasingly valuable for such analyses.

\acknowledgments
We thank Hiroyuki Nakano for careful reading of the manuscript and helpful comments.
This work was supported by JSPS KAKENHI Grants No. JP23H04891 (S. M.), No. JP23H04893 (S. M.), No. JP22K03639 (H. M.), and No. JP23KJ06945 (D. W.).

\section*{Data Availability}

The data that support the findings of this article are openly available~\cite{analysis_data}.

\appendix
\section{Offsets in the marginal posterior distributions} \label{append:offsets}

\begin{figure*}
    \centering
    \begin{minipage}{0.35\textwidth}
\includegraphics[width=\linewidth]{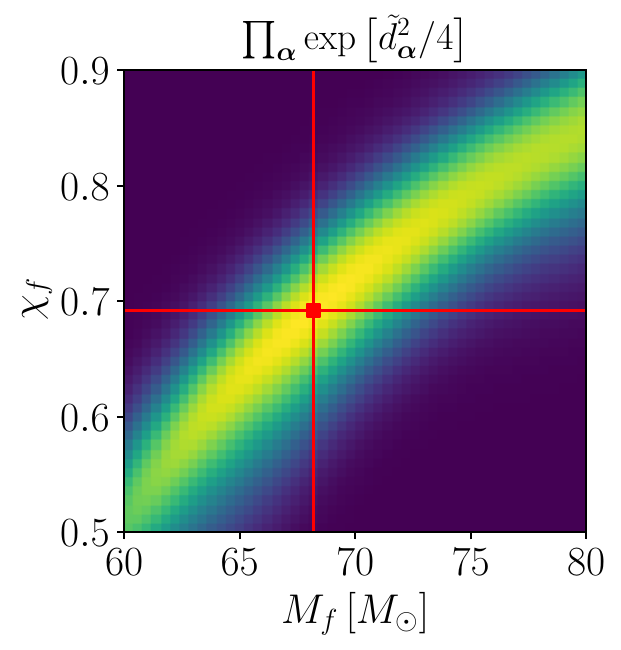}
    \end{minipage}
    \begin{minipage}{0.35\textwidth}
        \includegraphics[width=\linewidth]{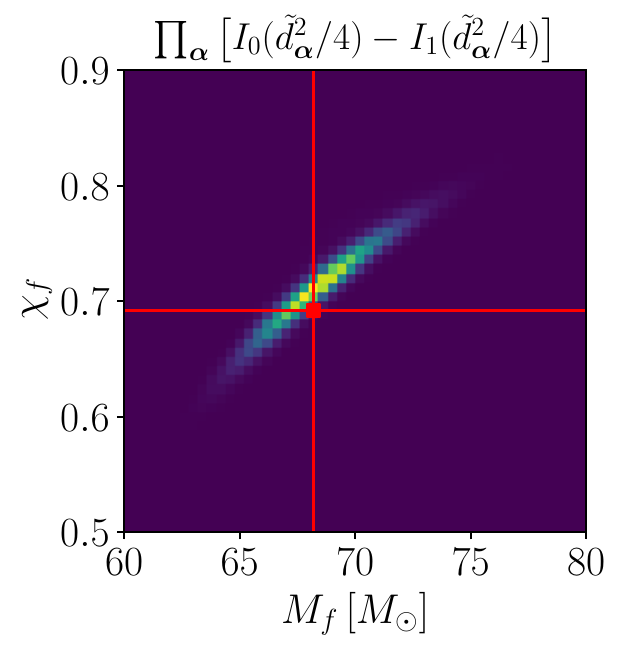}
    \end{minipage}
    \caption{
        Contributions of each term in the marginalized posterior distribution for $M_f$ and $\chi_f$ given in Eq. \eqref{eq:L_A_beta}.
        We consider the same settings of the eight-mode signal analysis in Sec. \ref{sec:damped}, i.e., we use the eight-mode template with $D=4$ to recover the eight-mode signal.
        The red lines represent the true values of $M_f$ and $\chi_f$: $M_f=68.2\,M_{\odot}$ and $\chi_f=0.692$.
    }
    \label{fig:likelihood_mass_spin}
\end{figure*}
\begin{figure*}
    \centering
    \begin{minipage}{0.4\textwidth}
\includegraphics[width=\linewidth]{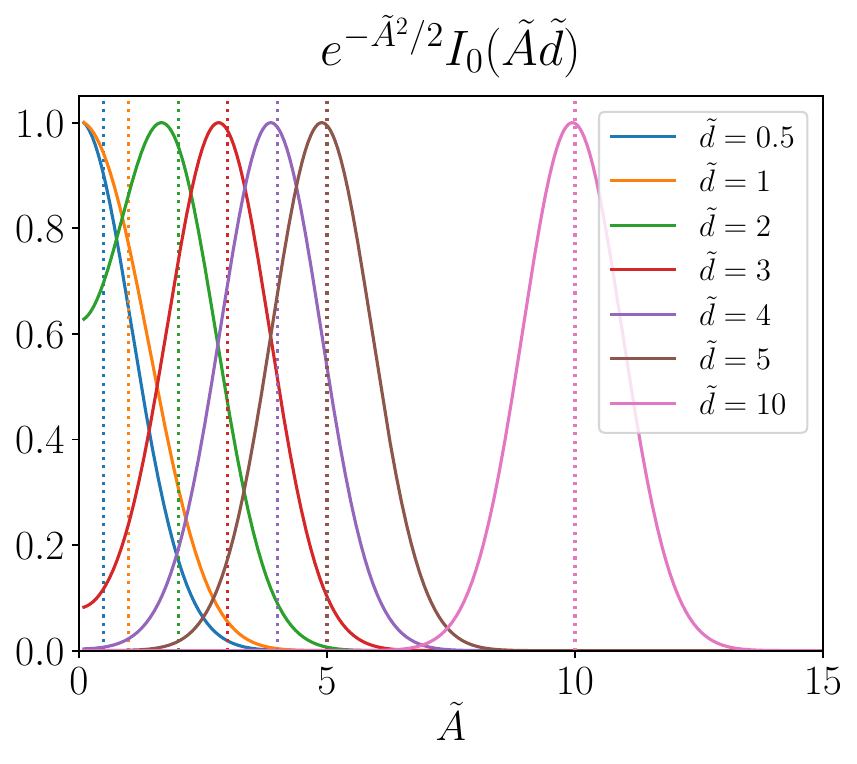}
    \end{minipage}
    \begin{minipage}{0.4\textwidth}
        \includegraphics[width=\linewidth]{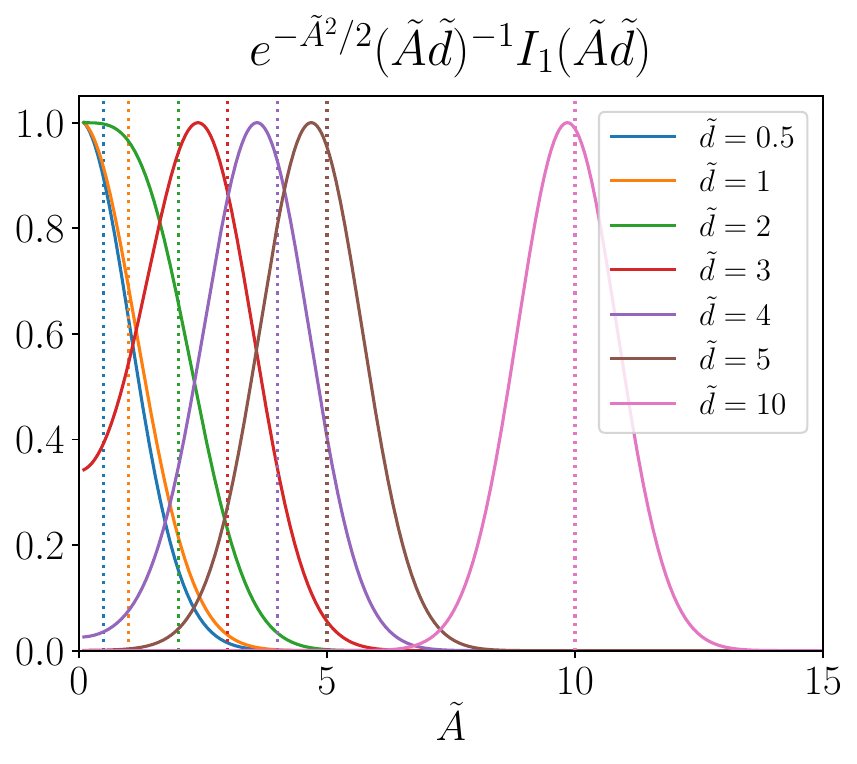}
    \end{minipage}
    \caption{
        Behavior of the marginal posterior distributions given in Eq. \eqref{eq:L_beta}.
        The left (right) panel shows the function in the case of $D=2$ ($D=4$).
        Dashed lines represent the values of $\tilde{d}$, which can be regarded as the true values of $\tilde{A}$ for each distribution.
    }
    \label{fig:likelihood_amplitude}
\end{figure*}
We examine the offset of the peak of the marginal posterior distribution for $\tilde{A}_{\bm{\alpha}}$ from the true value, noting that, in general, the peak of a marginal posterior does not necessarily coincide with the true value.

First, we consider the marginal posterior for $M_f$ and $\chi_f$ given in Eq.~\eqref{eq:L_A_beta}.
Figure \ref{fig:likelihood_mass_spin} shows the values of the exponential term $\prod_{\bm{\alpha}}\exp\normalsize[\tilde{d}^2_{\bm{\alpha}}/4\normalsize]$ and the Bessel function term $\prod_{\bm{\alpha}}[I_0(\tilde{d}^2_{\bm{\alpha}}/4)-I_1(\tilde{d}^2_{\bm{\alpha}}/4)]$ for $D=4$.
As the simulated data, we use the eight-mode signal described in Sec. \ref{sec:damped}, while for the template we assume the same set of \qnms with $D=4$, which corresponds to the same setting as in Fig.~\ref{fig:N=7_SNR=30}.
As seen in the figure, the exponential term peaks at the true value (shown in red), whereas the Bessel function term, arising from the analytic marginalization, does not peak at the true value.

Next, we consider the marginal posterior for $\tilde{A}_{\bm{\alpha}}$ given in Eq.~\eqref{eq:L_beta}.
The functional behavior of this marginal posterior is illustrated in Fig.~\ref{fig:likelihood_amplitude}.
For simplicity, we focus on the single-mode case, and the quantity $\tilde{d}$---determined by the data, $M_f$ and $\chi_f$---is set to the values indicated in the legend.
If the data contains only the signal, described by $\bm{d}=V(M_f^{\mathrm{true}}, \chi_f^{\mathrm{true}})\bm{c}^{\mathrm{true}}$, then the value of $\tilde{\bm{d}}$ corresponding to $M_f^{\mathrm{true}}$, $\chi_f^{\mathrm{true}}$ is given by
\begin{align}
    \tilde{\bm{d}}(M_f^{\mathrm{true}},\chi_f^{\mathrm{true}})
    &=\tilde{V}(M_f^{\mathrm{true}},\chi_f^{\mathrm{true}})R^{-1}\bm{d} \notag\\
    &=U^{-1}(M_f^{\mathrm{true}},\chi_f^{\mathrm{true}})\bm{c}^{\mathrm{true}} \notag\\
    &=\tilde{\bm{c}}^{\mathrm{true}},
\end{align}
where we used Eqs.~\eqref{eq:def_U}, \eqref{eq:orthonormality}, and \eqref{eq:amplitude_transformation}.
Therefore, the true value of $\tilde{d}_{\bm{\alpha}}$ corresponds to the true value of $\tilde{A}_{\bm{\alpha}}$.
The values of $\tilde{d}$, which can be regarded as the true values of $\tilde{A}$, are shown as dashed lines in Fig.~\ref{fig:likelihood_amplitude}.
We see that the marginal posterior exhibits an offset toward supporting values smaller than the true ones, particularly when the true value itself is small.
Moreover, this effect is more pronounced in the case of $D=4$.
This behavior is consistent with the offsets observed in Sec. \ref{sec:damped}.
This explains that the offset arises from the functional form of the analytically marginalized posterior, and does not lead to false positives.

\section{Additional tests}
\label{append:additional tests}

\begin{figure*}[htbp]
    \centering
    \begin{minipage}{0.45\textwidth}
        \includegraphics[width=\linewidth]{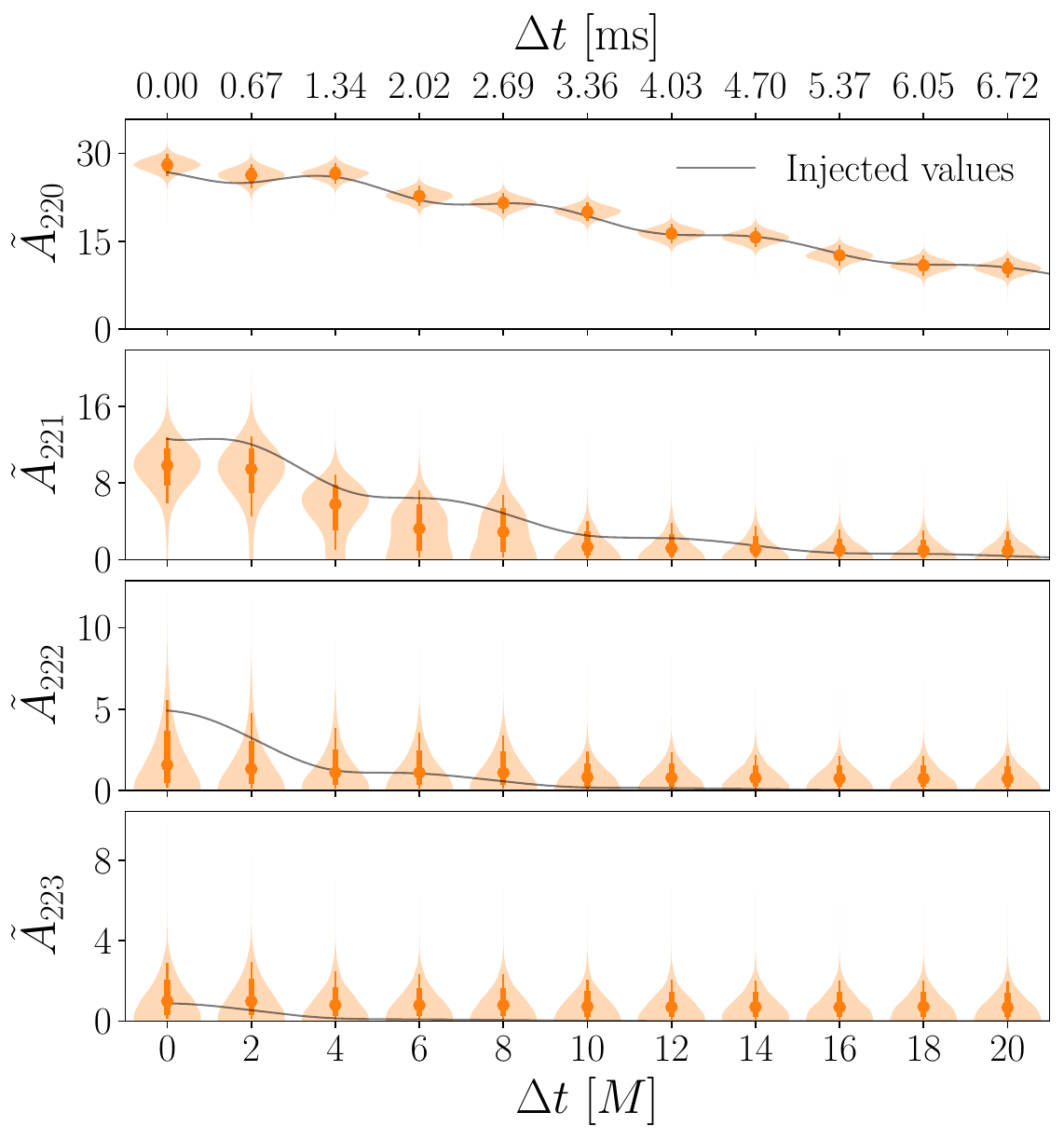}
    \end{minipage}
    \begin{minipage}{0.45\textwidth}
        \includegraphics[width=\linewidth]{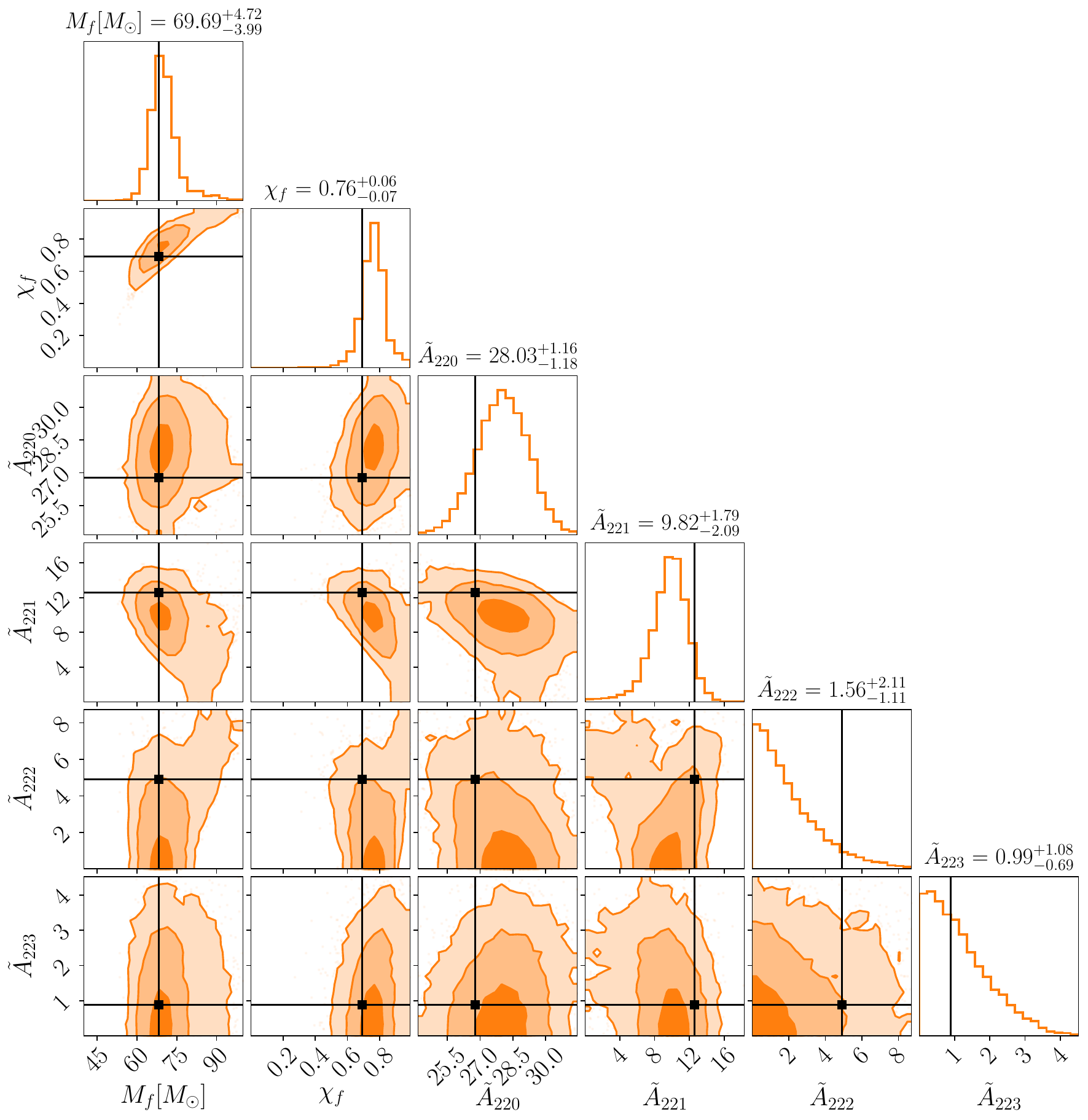}
    \end{minipage}
    \caption{Posterior distributions obtained by analyzing the eight-mode signal with the four-mode template.
    The left panel shows the results evaluated using data segments starting at different times.
    The right panel shows the results evaluated using a data segment starting at the peak time.}
    \label{fig:Kd=8_Kt=4}
\end{figure*}

\begin{figure*}[htbp]
    \centering
    \begin{minipage}{0.37\textwidth}
        \includegraphics[width=\linewidth]{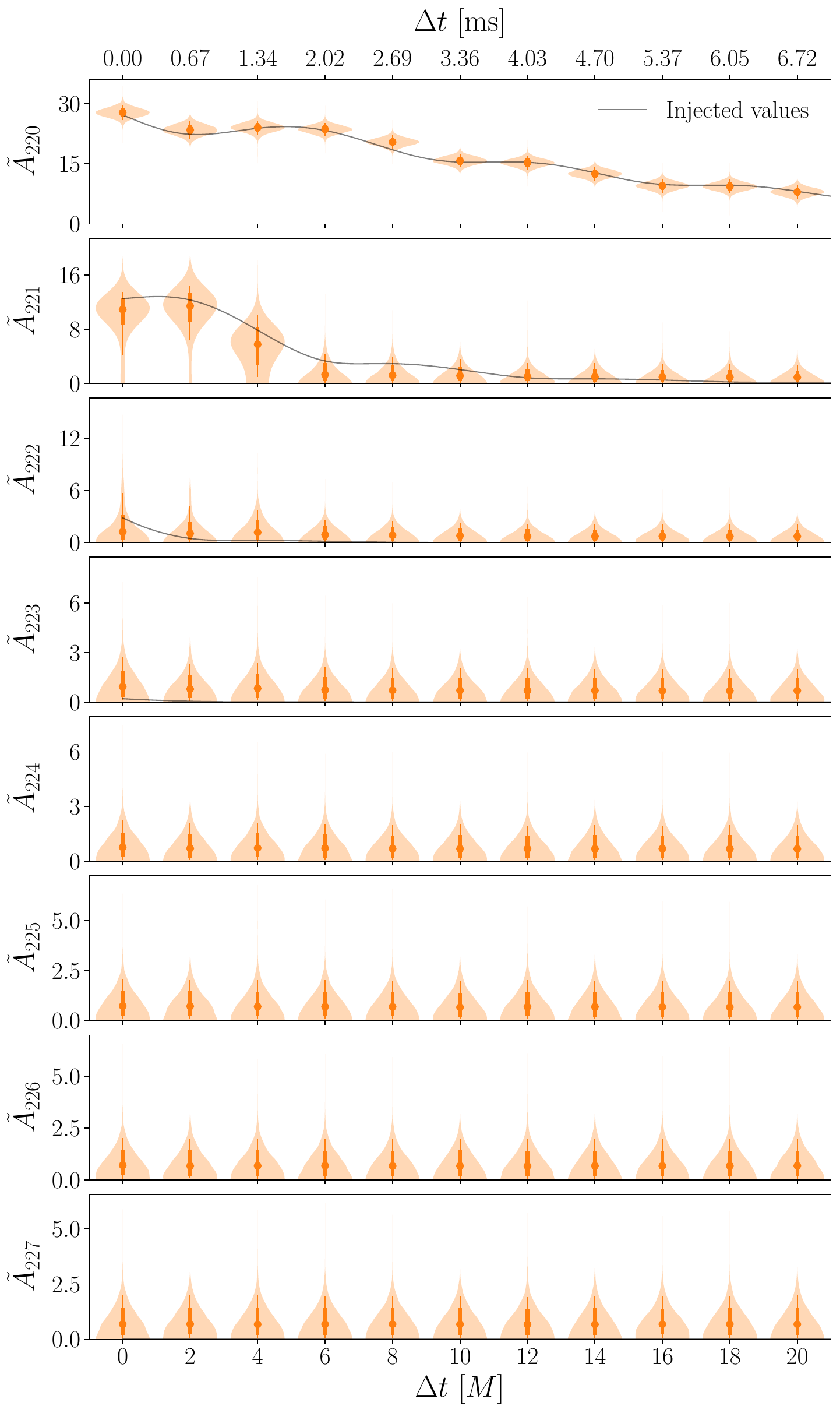}
    \end{minipage}
    \begin{minipage}{0.62\textwidth}
        \includegraphics[width=\linewidth]{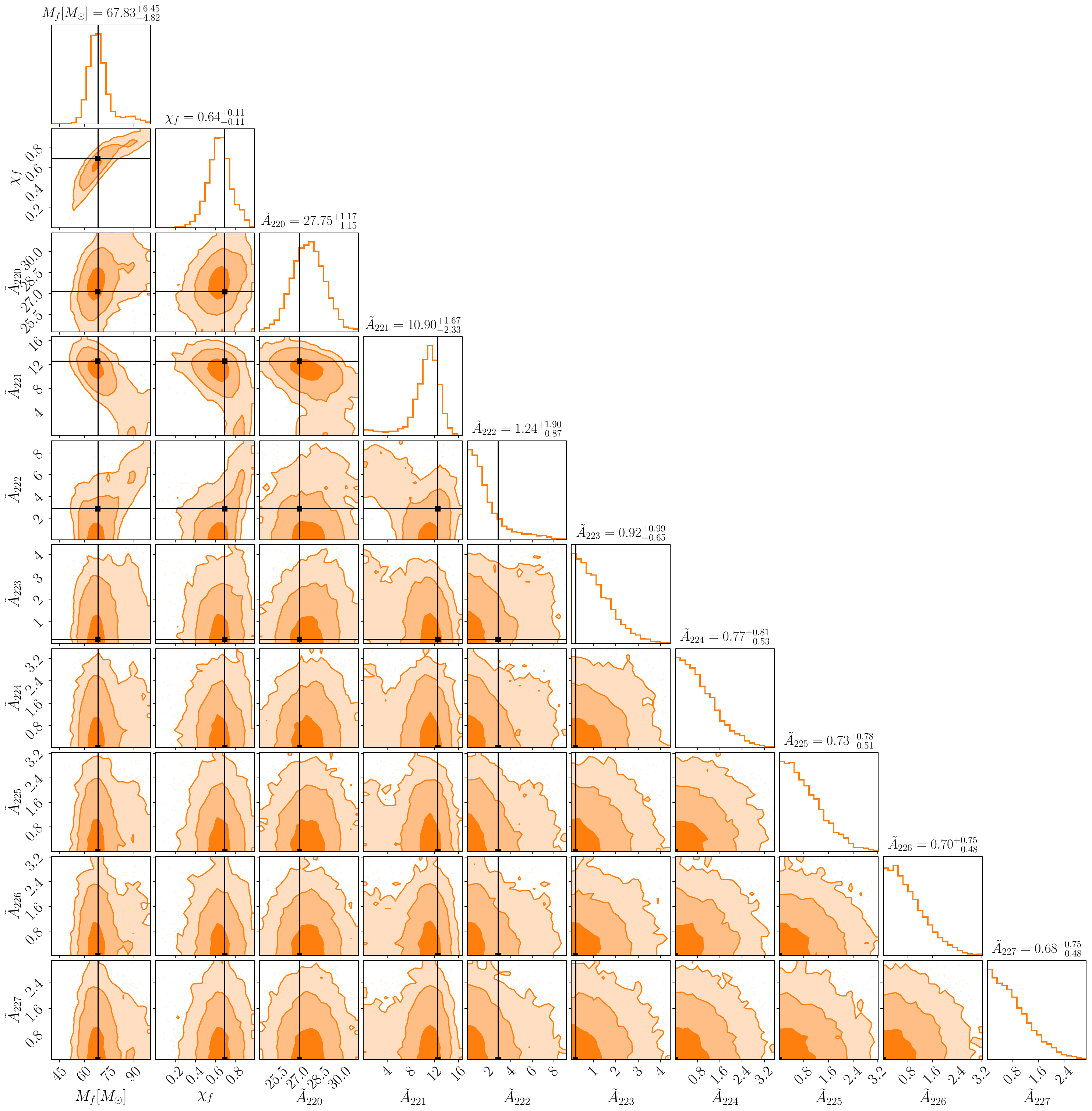}
    \end{minipage}
    \caption{Posterior distributions obtained by analyzing the four-mode signal with the eight-mode template.
    The left panel shows the results evaluated using data segments starting at different times.
    The right panel shows the results evaluated using a data segment starting at the peak time.}
    \label{fig:Kd=4_Kt=8}
\end{figure*}

In addition to the tests presented in Sec. \ref{sec:damped}, we also perform supplementary analyses in which the eight-mode signal is analyzed with the four-mode template, and conversely, the four-mode signal is analyzed with the eight-mode template.
The results are shown in Figs.~\ref{fig:Kd=8_Kt=4} and \ref{fig:Kd=4_Kt=8}, respectively.
These cases do not exhibit any false positives.

\bibliographystyle{apsrev4-1}

\end{document}